\UseRawInputEncoding

\documentclass[aip,jcp,amsmath,amssymb,floatfix,reprint,citeautoscript,noeprint,superscriptaddress,twocolumn, ]{revtex4-2}

\usepackage{dsfont}
\usepackage{lipsum} 
\usepackage{bibentry}
\usepackage[english]{babel}
\usepackage[dvipsnames]{xcolor}
\usepackage{graphicx}
\usepackage[caption=false]{subfig} 
\usepackage{amsmath,amssymb,bm}
\usepackage[version=3]{mhchem}
\usepackage{verbatim}
\usepackage{multirow}
\usepackage{dcolumn}
\usepackage{float}
\usepackage{nicefrac}
\usepackage{siunitx}
\usepackage{booktabs}
\usepackage{chemformula}
\usepackage{wrapfig}
\usepackage{enumitem}  
\usepackage{transparent}
\usepackage[colorlinks,allcolors=black,citecolor=blue,urlcolor=blue]{hyperref}
\usepackage{tcolorbox}
\usepackage{relsize}

\hypersetup{
    citecolor=Turquoise,
    colorlinks=true,
    linkcolor=black,    
    urlcolor=Turquoise,
    }
\DeclareSIUnit[number-unit-product = {\,}]{\amu}{amu}
\DeclareSIUnit[number-unit-product = {\,}]{\kJmol}{\kilo\joule\per\mol}
\DeclareSIUnit[number-unit-product = {\,}]{\Nsm}{\newton\second\per\meter\cubed}
\DeclareSIUnit[number-unit-product = {\,}]{\THz}{\tera\hertz}
\DeclareSIUnit[number-unit-product = {\,}]{\meV}{\milli\electronvolt}
\DeclareSIUnit[number-unit-product = {\,}]{\cal}{cal}

\newcommand{\mbf}[1]{\boldsymbol{\mathit{#1}}}
\newcommand{\mrm}[1]{\mathrm{#1}}
\newcommand{\mcl}[1]{\mathcal{#1}}
\newcommand{\tcr}[1]{\textcolor{black}{#1}}

\usepackage{sansmathfonts}
\usepackage[justification=centerlast, font={sf}, labelfont={bf}]{caption}
\begin{document}


\title{\tcr{Revisiting the Green--Kubo relation for friction in nanofluidics}}

\author{Anna T. Bui}
\affiliation{Yusuf Hamied Department of Chemistry, University of
  Cambridge, Lensfield Road, Cambridge, CB2 1EW, United Kingdom}

\author{Stephen J. Cox}
\email{stephen.j.cox@durham.ac.uk}
\affiliation{Yusuf Hamied Department of Chemistry, University of
  Cambridge, Lensfield Road, Cambridge, CB2 1EW, United Kingdom}
\affiliation{Department of Chemistry, Durham University, South Road, Durham, DH1 3LE, United Kingdom}

\date{\today}

\color{darkgray}

\begin{abstract}
A central aim of statistical mechanics is to establish connections
between a system's microscopic fluctuations and its macroscopic
response to a perturbation. For non-equilibrium transport properties,
this amounts to establishing Green--Kubo (GK) relationships. In
hydrodynamics, relating such GK expressions for liquid--solid friction
to macroscopic slip boundary conditions has remained a long-standing
problem due to two challenges: (i) The GK running integral of the
force autocorrelation function decays to zero rather than reaching a
well-defined plateau value; and (ii) debates persist on whether such a
transport coefficient measures an intrinsic interfacial friction or an
effective friction in the system. Inspired by ideas from the
coarse-graining community, we derive a GK relation for liquid--solid
friction where the force autocorrelation is sampled with a constraint
of momentum conservation in the liquid.  Our expression does not
suffer from the ``plateau problem'' and unambiguously measures an
effective friction coefficient, in an analogous manner to Stokes'
law. We further establish a link between the derived friction
coefficient and the hydrodynamic slip length, enabling a
straightforward assessment of continuum hydrodynamics across length
scales.  We find that continuum hydrodynamics describes the simulation
results quantitatively for confinement length \tcr{scales} all the way
down to \tcr{1\,nm.  Our approach} amounts to a straightforward
modification to the present standard method of quantifying interfacial
friction from molecular simulations, making possible a sensible
comparison between surfaces of vastly different slippage.
\end{abstract}

\maketitle


\section{Introduction}

Describing the flow of liquids at solid surfaces is essential for 
understanding many physical processes of both fundamental and 
technological importance, including transport through membranes 
and nanopores, power generation, desalination and electrokinetic 
effects \cite{Siria2013,Joly2021, Bocquet2020}.
Unlike in the bulk, fluid transport under confinement is 
governed by frictional forces arising from momentum transfer 
at the liquid--solid interface. Over the past decade, 
advances in device fabrication \cite{Geim2013,Celebi2014,Bhardwaj2024} 
have spurred increased interest in nano-confined water,
with many experimental and simulation studies reporting
exotic friction effects in ``one-dimensional'' (1D) nanotubes 
and ``two-dimensional''(2D) nanochannels \cite{Holt2006,Majumder2005,Whitby2008,Secchi2016,Tunuguntla2017,Keerthi2021, Ma2016, Falk2010, Thomas2008,
Tocci2014, Thiemann2022, Hummer2001,Kalra2003, Bui2023,Striolo2006}.
A natural consequence of confinement is that as the 
confining volume is decreased, the surface-to-volume ratio increases, 
which amplifies the impact of interfacial effects on transport
properties like friction. In addition to dissipation 
processes at the interface, viscous effects in the liquid itself also
contribute to the effective friction in the system.
Therefore, it is essential to distinguish these competing
effects by quantifying an intrinsic surface property 
that is independent of the confining volume.

From a macroscopic perspective, such a property is the slip 
length $b$, measuring the distance beyond the surface of 
the wall at which the fluid velocity $v_x(z)$ extrapolates to zero.
The slip length enters a continuum hydrodynamics description,
hereafter referred to as ``classical hydrodynamic theory''
(CHT), as \tcr{a boundary condition to the Navier--Stokes equations}
\begin{equation}
    \frac{\partial v_x}{\partial z}\bigg|_{z=z_0} = \frac{v_{\mrm{s}}}{b},
    \label{eq:boundary}
\end{equation}
relating the velocity gradient with the slip velocity $v_{\mrm{s}}=v_x(z_0)$.
\tcr{The $z=z_0$ plane is where the hydrodynamic boundary position is placed.}
The no-slip boundary condition $b=0$ applies for surfaces that are
highly sticky, while for atomically smooth and non-wetting surfaces
where there is finite slippage at the interface, $b>0$.  The gradient
of the fluid velocity is linearly related to the viscous stress of the
fluid $\sigma_{xz}=-\eta(\partial v_x/\partial z)$, where $\eta$ is
the shear viscosity \tcr{(which we assume takes its bulk value in all
regions occupied by the fluid)}. The stress is itself balanced by the
friction force from the solid to the liquid $F_x$ per unit surface
area $\mcl{A}$.  This allows a linear relation between $F_x$ and
$v_{\mrm{s}}$ to be written
\begin{equation}
    F_x = -\lambda_{\mrm{intr}} \mcl{A} v_{\mrm{s}}.
    \label{eqn:navier}
\end{equation}
Eq.~\ref{eqn:navier} is known as Navier's interfacial
constitutive relation\cite{Navier1823}, defining an intrinsic friction coefficient $\lambda_{\mrm{intr}}$.
Both $\lambda_{\mrm{intr}}$ and $b$ are intrinsic properties
of the interface  and are related by
\begin{equation}
    \lambda_{\mrm{intr}} = \frac{\eta}{b}.
    \label{eqn:intrinsic_friction}
\end{equation}
It has been established that in the linear response regime,
$b$ (and therefore $\lambda_{\mrm{intr}}$) is independent of both
the type of flow (i.e., Couette or Poiseuille) and the channel 
height \cite{Cieplak2001}, provided that 
the length scale of confinement is large enough for CHT to faithfully describe hydrodynamic transport. 
Quantifying ``large enough'' is something that the framework we present in this article allows us 
to directly assess with molecular simulations; generally speaking, we find CHT works with as few as two or three layers of water.

In experiments\cite{Holt2006, Secchi2016, Keerthi2021,Whitby2008, Majumder2005}, 
to measure transport coefficients such as $\lambda_{\rm intr}$,
one would drive the system of interest
out of equilibrium and measure the responding flux to the 
driving force. The resulting force--flux relation obtained can 
then be mapped back onto predictions from CHT, and $b$ or 
$\lambda_{\mrm{intr}}$ can be backed out accordingly.
While an analogous strategy can in principle be followed in molecular simulations, 
methods using non-equilibrium molecular dynamics (NEMD) are limited by statistical sampling and delicate issues that arise in thermostatting the system \cite{Kannam2012}.

From a microscopic perspective, frictional effects coming 
from dissipation in the liquid manifest through balance 
of forces at non-equilibrium steady states. 
A natural question to ask is if one can relate 
these transport coefficients to equilibrium fluctuations 
in the liquid, invoking Onsager's regression hypothesis. 
In this context, one attempts to seek Green--Kubo (GK)
expressions for
a liquid--solid friction coefficient, allowing 
the hydrodynamic boundary conditions to be characterized from 
a single equilibrium molecular dynamics (EMD) simulation.
The seminal work by Bocquet and Barrat\cite{Bocquet1994, Bocquet2013} 
(BB) proposed such a relation for $\lambda_{\rm intr}$, given as
\begin{equation}
   \lambda_{\mrm{BB}} = \frac{\beta}{\mcl{A}} \int^\infty_0\!\mrm{d}t\,\langle F_x(t) F_x(0) \rangle,
    \label{eqn:BB}
\end{equation}
where $\beta=1/(k_{\mrm{B}}T)$, $k_{\mrm{B}}$ is the Boltzmann constant,
$T$ is temperature and $\langle\ldots\rangle$ indicates an ensemble
average at equilibrium. Such a GK relation involving 
the force autocorrelation function is similar to the 
expression for friction $\xi$ of a heavy Brownian particle
immersed in a bath  of lighter particles, as derived from Langevin dynamics. 
In Ref.~\onlinecite{Bocquet2013}, BB presented a derivation for
Eq.~\ref{eqn:BB}  from a Langevin equation for the stochastic
motion of the solid wall immersed in the fluid.

However, two major challenges arise upon application of
Eq.~\ref{eqn:BB} to measure friction at the liquid--solid interface. 
The first issue is the well-documented ``plateau problem,''
\cite{Kirkwood1946, Mazur1970, Espanol1993, Bocquet1997,
toda2012statistical, delaTorre2019, Oga2019, Oga2023}
referring to the fact that the integral in Eq.~\ref{eqn:BB} 
in general will decay to zero at long time. While this behavior is often attributed to the 
swapping of the order in which the thermodynamic and long-time limits are taken
\cite{Espanol1993, Oga2019, Chen2015, Bocquet1997, Bocquet2013},
we will show---as becomes clear when employing a Mori--Zwanzig formalism to derive the appropriate Langevin dynamics---that
this is in fact a direct consequence of using real forces 
rather than projected forces in Eq.~\ref{eqn:BB}. The plateau problem is most severe when the surface
is strongly wetting or when the liquid film is very 
thin (see \tcr{Fig.~S2 of the supplementary material}). 

The second major challenge involves the intense debate 
whether $\lambda_{\mrm{BB}}$ provides a measure of the 
intrinsic friction coefficient $\lambda_{\mrm{intr}}$ 
or the effective friction of the entire system
\cite{Petravic2007, Hansen2011, Huang2014, Chen2015,
Ramos-Alvarado2016, Nakano2020, Varghese2021}. 
From the perspective of Langevin theory, 
the friction coefficient $\xi$ characterizes the total
dissipation associated with the motion of the Brownian particle 
through the solvent, rather than just the intrinsic slippage at
the particle's surface. Therefore, directly relating
Eq.~\ref{eqn:BB} to the intrinsic friction $\lambda_{\mrm{intr}}$ 
is not straightforward.
These issues have thus made it difficult 
to faithfully assess friction under confinement 
across different length scales using Eq.~\ref{eqn:BB}. 
In this context, 
we note of the work of Petravic and Harrowell, who pointed 
out that for Couette flow, the friction measured by $\lambda_{\mrm{BB}}$
includes contributions both from the slip at the interface
and viscous dissipation in the fluid \cite{Petravic2007}.
Nonetheless, the GK relation given by Eq.~\ref{eqn:BB}
continues to be widely applied in the community 
\cite{Li2024,Liang2024, Tocci2020, Tocci2014, Bui2023,
Thiemann2022, KumarVerma2022, Seal2021,Joly2016}.
However, its use is often limited to surfaces of 
high slippage. There is thus a significant need to establish both 
a practical solution to the plateau problem, and a rigorous connection between the GK transport coefficient
and the intrinsic friction defined by CHT.

In this paper, we present a derivation of a GK 
relation for liquid--solid friction by considering the
liquid's stochastic motion in the solid's frame of reference,
using the Mori--Zwanzig projection operator formalism and 
linear response theory. The resulting expression for the friction coefficient, which overcomes the plateau problem, 
involves sampling the force autocorrelation function with a simple
momentum conservation constraint on the liquid.
We also show that this friction coefficient 
measures an effective friction of the system rather than
the intrinsic friction. By making an appropriate mapping to CHT, we 
then nonetheless relate $\lambda_{\rm eff}$ directly to the hydrodynamic
slip length $b$. We use the resulting framework 
to faithfully assess interfacial friction arising from water flow in 1D tubes and 
2D channels across a wide range of confining length scales, including results from first-principles-level simulations.

\section{The Green--Kubo relation for friction}

\subsection{Mori--Zwanzig projection operator formalism}

At the heart of any friction problem is a separation of 
timescales for different degrees of freedom.
In the textbook problem of Brownian motion, the two 
timescales at play are evident: 
one associated with the slow motion of the heavy Brownian
particle and one with its frequent collisions with lighter 
solvent particles, which constitute ``the bath.''
Instead of describing the complex system as a whole,
one tends to focus only on the motion of the Brownian particle as 
a coarse-grained variable, with the rapid collisions with the solvent treated 
as a fluctuating random force.
This approach underpins the Langevin equation, which describes 
the Brownian particle's motion using a combination
of a frictional drag force and a fluctuating random force.
More generally, this description can also be applied to an
arbitrary coarse-grained variable in any complex system,
provided that the variable evolves on a slower timescale 
than the rest of the system. This is an important result 
of Mori--Zwanzig theory\cite{Zwanzig1961,Mori1965}, where
projection operators are used to derive generalized Langevin 
equations for coarse-grained variables. Detailed presentation 
of Mori--Zwanzig theory can be found in standard texts \cite{ZwanzigBook,bernePecoraBook,HansenMcDonaldBook},
and here we will review only the most salient aspects for the job at hand.

For any phase variable $A$ evolving under equations of
motion from a specified Hamiltonian $\mcl{H}$, 
its time evolution obeys the Liouville equation 
\begin{equation}
    \frac{\mrm{d}A(t)}{\mrm{d}t} = i\mcl{L}A(t),
\end{equation}
which has the formal solution $A(t)=\mrm{e}^{i\mcl{L}t}A(0)$, 
where $i\mcl{L}$ denotes the Liouville operator associated with $\mcl{H}$. 
For any observable $B$, we can denote its projection onto $A$ 
with a projection operator $\mcl{P}$
\begin{equation}
    \mcl{P}B(t) = (B(t),A) (A,A)^{-1} A,
\end{equation}
where $(\cdots,\cdots)$ denotes a scalar product,
\begin{equation}
(B(t),A) = \int\!\mrm{d}\Gamma\,f_0(\Gamma)B(\Gamma, t)A^\ast(\Gamma),
\end{equation}
with $f_0(\Gamma)$ the equilibrium distribution of initial phase space points $\Gamma$.
The complementary operator $\mcl{Q}=\mathds{1}-\mcl{P}$ projects onto 
the subspace orthogonal to $A$, such that
\begin{equation}
    (\mcl{Q} B(t), A) = 0.
\end{equation}

The time evolution of the stochastic coarse-grained variable $A$ is given by the generalized Langevin equation,
\begin{equation}
     \frac{\mrm{d}A(t)}{\mrm{d}t} = i\varOmega A(t) -\int^t_0\!\mrm{d}t^\prime\, K(t-t^\prime) A(\tcr{t^\prime}) + F^{\mrm{R}}(t),
     \label{eqn:GLE}
\end{equation}
where $i\varOmega$ (``the frequency''), $K$ (``the memory function'') and $F^{\mrm{R}}$ (``the random force'') are well-defined
functions. 
When $A$ is a 
single coarse-grained variable, $i\varOmega=0$.
The memory function is related to the autocorrelation 
of the random force
\begin{equation}
  K(t) = (F^{\mrm{R}}(t), F^{\mrm{R}}) (A, A)^{-1},
  \label{eqn:memoryfunction}
\end{equation}
while the random force itself is given by 
the projection of $\dot{A}$ orthogonal to $A$
\begin{equation}
     F^{\mrm{R}}(t)=\mrm{e}^{i\mcl{QL}t}F^{\mrm{R}}(0),
    \label{eqn:randomforce}
\end{equation}
where $F^{\mrm{R}}(0) = \mcl{Q}\dot{A}(0)$.
An important consequence of the orthogonal projection
in the random force is the lack of correlation
between $ F^{\mrm{R}}$ and $A$, i.e
\begin{equation}
   (F^{\mrm{R}}(t), A)  = 0,
    \label{eqn:uncorrelated}
\end{equation}
which is at the foundation of the Onsager regression hypothesis. 

When $A$ relaxes much more slowly than the random noise, which implicitly assumes 
it is the only slow degree of freedom, the memory function can be treated 
in the Markovian approximation: $K(t)=2\Lambda \delta(t)$. The generalized Langevin equation (Eq.~\ref{eqn:GLE}) 
then simplifies to the Langevin equation,
\begin{equation}
     \frac{\mrm{d}A(t)}{\mrm{d}t} = -\varLambda  A(t) + F^{\mrm{R}}(t), 
     \label{eqn:Markov}
\end{equation}
where a time-independent friction coefficient is
given as
\begin{equation}
   \varLambda = \int^\infty_0\!\mrm{d}t\,(\mrm{e}^{i\mcl{QL}t}
 F^{\mrm{R}},F^{\mrm{R}})  (A,A)^{-1}.
 \label{eqn:friction_projected}
\end{equation} 
%

\subsection{Applying the operator formalism to liquid--solid friction}

\begin{figure}[t]
  \includegraphics[width=\linewidth]{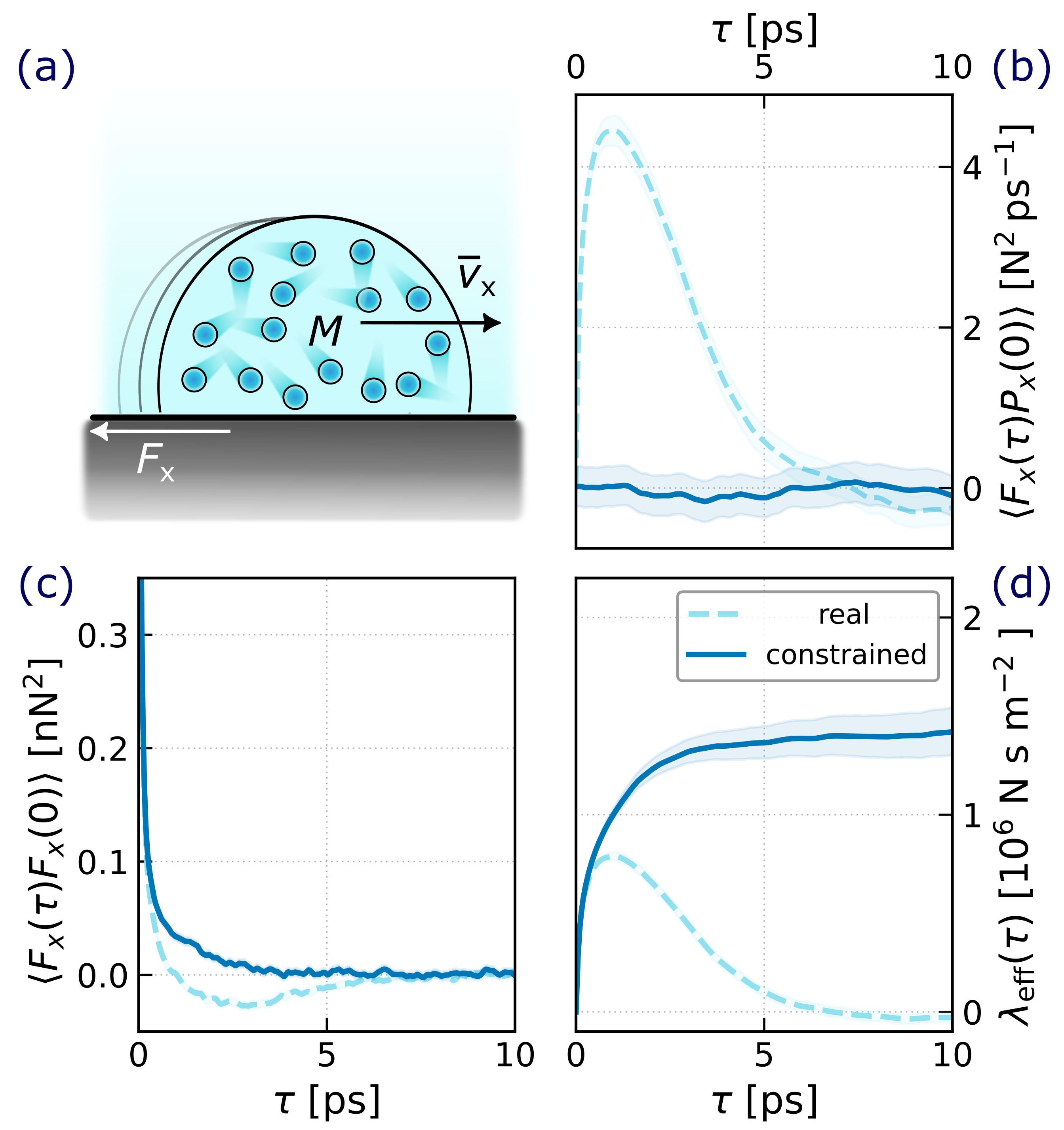}
  \caption{\textbf{The Green--Kubo friction integral.} \tcr{(a) Schematic of the system set up that we use for the derivation; a liquid droplet of total mass $M$ moves stochastically on a solid surface. Its center-of-mass velocity along one direction is indicated by $\bar{v}_x$, and the frictional force with the surface is $F_x$.} For
  a liquid--solid interface with high wettability (see Sec.~\ref{sec:Interpreting::Connection}), we show \tcr{(b)} the
  force--momentum correlation function, \tcr{(c)} force--force
  autocorrelation function and \tcr{(d)} the friction GK integral.
Using real dynamics results in a plateau problem
since the force is correlated with the momentum, 
leading to negative correlation in the  force autocorrelation; this causes the GK integral to vanish at long times.
Constrained dynamics \tcr{bypasses} this issue by \tcr{imitating the effect} of projected forces, ensuring that the integral plateaus \tcr{to} a finite value.}
\label{fig1}
\end{figure}

To describe liquid--solid friction, we will consider 
the problem of a liquid droplet's stochastic motion on 
a solid surface, schematically shown in Fig.~\ref{fig1}(a).
\tcr{We choose this arrangement both for conceptual simplicity, and to make
the connection between this derivation and that for Brownian motion
clear; the resulting framework, however, is readily applied to systems
comprising a fluid confined between walls.} We will work in the
solid's frame of reference so that we can consider it
to be at rest.  The liquid droplet comprises $N$ particles, whose
positions and momenta are $\{\mbf{r}_i,\mbf{p}_i\}$ and has a total
mass of $M=\sum^N_{i=1} m_i$. For simplicity, we will focus only on
the droplet's motion in the $x$ direction, with a center-of-mass
velocity of $\overline{v}_x$.

Provided that the droplet is sufficiently large, 
two very different timescales can be identified in 
this problem: one associated with the slow motion of the 
droplet undergoing a random walk on the surface and one associated 
with frequent collisions between individual liquid particles 
and the solid. Analogous to Brownian motion, 
we therefore now treat the total linear momentum of 
the liquid droplet $P_x=M\overline{v}_x$ as the 
coarse-grained variable of interest. Its time 
evolution is governed by
\begin{equation}
    \frac{\mrm{d}P_x(t)}{\mrm{d}t} = i\mcl{L}P_x(t), \quad\quad\mrm{[real\,dynamics]}
\end{equation}
for which the solution is $P_x(t)=\mrm{e}^{i\mcl{L}t}P_x(0)$ 
where the Liouville operator is explicitly given as  
\begin{equation}
    i\mcl{L} = \sum^N_{i=1}\left( \frac{\mbf{p}_i}{m_i}\cdot \frac{\partial }{\partial \mbf{r}_i} + \mbf{f}_i\cdot \frac{\partial }{\partial \mbf{p}_i}\right),
\end{equation}
where $\mbf{f}_i$ is the force on each liquid particle.
We will refer to the dynamics associated with $i\mcl{L}$ as 
the ``real dynamics,'' i.e., those that result from propagating
Newton's equations of motion in standard MD simulations.
Upon application of the result in Eq.~\ref{eqn:Markov},
which implies a Markovian approximation, the liquid motion 
can be described with
\begin{equation}
    M \frac{\mrm{d}\overline{v}_x(t)}{\mrm{d}t} = -\lambda_{\mrm{eff}}\mcl{A}\overline{v}_x(t) + F^{\mrm{R}} (t),
    \label{eqn:Markov-droplet}
\end{equation}
where $\lambda_{\rm eff}$ is an effective friction coefficient. 
The term $-\lambda_{\mrm{eff}}\mcl{A}\overline{v}_x(t)$
measures the frictional drag force on the liquid droplet as 
a whole, rather than the intrinsic friction due to slippage
at the interface $-\lambda_{\mrm{intr}}\mcl{A}\overline{v}_\mrm{s}$
as in Eq.~\ref{eqn:navier} (see Sec.~\ref{sec:Interpreting::Effective}).
The random force is now given by the orthogonal projection of
total tangential force on the liquid  
\begin{equation}
    F^{\mrm{R}}(t)=\mrm{e}^{i\mcl{QL}t}F^{\mrm{R}}(0), 
\end{equation}
where $F^{\rm R}(0)=\dot{P}_x(0)=F_x(0)$.
Using the fluctuation--dissipation theorem 
as stated in Eq.~\ref{eqn:friction_projected} and
the equipartition theorem $M\langle \overline{v}_x^2 \rangle=k_{\mrm{B}}T$, the friction coefficient is given as 
\begin{equation}
    \lambda_{\rm eff} = \frac{\beta}{\mcl{A}}\int^\infty_0\!\mrm{d}t\,( \mrm{e}^{i\mcl{QL}t}  F_x, F_x) \\
    \label{eqn:lambda_projected}
\end{equation}

According to Eq.~\ref{eqn:lambda_projected}, 
to obtain $\lambda_{\rm eff}$, the force autocorrelation 
function should be sampled with projected dynamics
propagated by $i\mcl{QL}$. However, such a dynamical scheme is not readily realizable in MD simulations, 
since in general it is not possible write the projection operator $\mathcal{Q}$ explicitly.
To make progress, a common though ad-hoc
assumption to make is that the random force evolves
with $i\mcl{L}$ instead of $i\mcl{QL}$, i.e
$F^{\mrm{R}}(t)\approx\mrm{e}^{i\mcl{L}t}F_x(0)$. 
The BB formula given by Eq.~\ref{eqn:BB} 
amounts to such an approximation.
Though the GK formula in Eq.~\ref{eqn:BB} can be
sampled directly in a MD simulation, 
a consequence of replacing $i\mcl{Q}\mcl{L}$ with $i\mcl{L}$,
is that the random force is no longer uncorrelated 
with the coarse-grained variable $P_x$, 
violating the condition stated in Eq.~\ref{eqn:uncorrelated}.
The effect is easily seen in the results of MD simulations. For a water film $\sim 2.7\,$nm in thickness 
in contact with a strongly wetting surface (see Sec.~\ref{sec:Interpreting::Connection}), we see in Fig.~\ref{fig1}(b) 
that $\langle F_x(\tau)P_x(0)\rangle$ initially shows a large positive correlation, 
exhibiting a weak negative correlation at longer times.
Such correlation between the force and the momentum
leads to a negative contribution to the force 
autocorrelation function, as shown in 
Fig.~\ref{fig1}(c), making the integral in Eq.~\ref{eqn:BB} vanish
at long times, as seen in Fig.~\ref{fig1}(d).
We can also understand why the plateau problem encountered
in the use of the BB formula is most severe for surfaces of 
strong wettability or \tcr{when the liquid film is very thin}.
In these cases, since a large portion of momentum in the 
liquid can be transfered to the solid, $F_x$ and $P_x$ 
are strongly correlated in the real dynamics, and 
Eq.~\ref{eqn:uncorrelated} is severely violated.
Moreover, the Markovian approximation
that relaxation of $P_x$ is much slower than 
all the other degrees of freedom also becomes more severe.

\subsection{Momentum conservation constraint to overcome the plateau problem}

In fact, the plateau problem has been discussed in 
the coarse-graining community, where the 
issues associated with replacing projected
dynamics with real dynamics, 
along with the limitations of the Markovian approximation,
are more widely recognized \cite{Hijon2010,Espanol2019}.
There have also been developments of algorithms 
for obtaining projected observables of the GLE directly \cite{Carof2014, Ayaz2022, Dominic2023}.
Here, we adopt the strategy proposed in Ref.~\onlinecite{Hijon2010}:
instead of modeling the real dynamics directly, where the
Markovian approximation is only good in certain limits, 
we aim to construct a constrained dynamics in which
the Markovian approximation is enforced. 
It then becomes a modeling question of how well 
the constrained dynamics represents the physical
behavior of interest. In the context of 
liquid--solid friction, we validate our results 
directly against NEMD simulations.

To make such a Markovian approximation exact,
we want to ensure a separation of timescales
between the motion of the droplet and the other
degrees of freedom. To this end, 
we introduce a constraint on the system that maintains
the droplet's linear momentum to be zero at all times.
In this picture, the motion of the droplet as a whole
is infinitely slow compared to the individual
microscopic processes that each fluid particle undergoes.
Since $P_x$ does not evolve in time under the constraint, 
the Liouville equation is trivially
\begin{equation}
    \frac{\mrm{d}P_x(t)}{\mrm{d}t} = i\mcl{L}_{\mrm{c}}P_x(t)=0, \quad\quad\mrm{[constrained\,dynamics]}
\end{equation}
such that $P_x(t)=\mrm{e}^{i\mcl{L}_{\mrm{c}}t}P_x(0)=P_x(0)=0$. 
The corresponding Liouville operator to fix the momentum of the 
liquid can be written as
\begin{equation}
    i\mcl{L}_{\mrm{c}} = \sum^N_{i=1}\left[ \left(\frac{\mbf{p}_i}{m_i} - \mu\hat{\mbf{e}}_x\right)\cdot \frac{\partial }{\partial \mbf{r}_i} + \left(\mbf{f}_i - \gamma m_i\hat{\mbf{e}}_x\right)\cdot \frac{\partial }{\partial \mbf{p}_i}\right],
\end{equation}
where $\hat{\mbf{e}}_x$ denotes the unit vector in the $x$ 
direction and the Lagrange multipliers are $\mu=P_x/M$ 
and $\gamma=F_x/M$. Such a constrained dynamics can be straightforwardly
realized in an MD simulation: at equilibrium, the liquid has 
zero initial momentum so the constraint should maintain $P_x=0$, which 
amounts to simply subtracting the center-of-mass velocity 
of the liquid at every time-step.

An important consequence of the constrained dynamics
is that since the liquid momentum $P_x$ is conserved,
all other degrees of freedom of the system 
become uncorrelated with $P_x$.
In other words, they all lie on the subspace orthogonal 
to $P_x$ such that the effect of the projection
operator $\mcl{Q}$ is already encapsulated, i.e., 
\begin{equation}
    \mcl{Q}i\mcl{L}_{\mrm{c}} = (\mathds{1}-\mcl{P})i\mcl{L}_{\mrm{c}} = i\mcl{L}_{\mrm{c}}.
\end{equation}
The random force is therefore now exactly
the force sampled with constrained dynamics
\begin{equation}
 F^{\mrm{R}}(t)=\mrm{e}^{i\mcl{L}_{\mrm{c}}t}F^{\mrm{R}}(0),
\end{equation}
where $F^{\rm R}(0)=\dot{P}_x(0)=F_x(0)$.
As a result, we can justifiably replace $i\mcl{QL}$ with $i\mcl{L}_{\mrm{c}}$ in Eq.~\ref{eqn:lambda_projected},
giving an expression for the friction coefficient 
\begin{equation}
\begin{split}
    \lambda_{\mrm{eff}} & = \frac{\beta}{\mcl{A}}\int^\infty_0\!\mrm{d}t\,(\mrm{e}^{i\mcl{L}_{\mrm{c}}t}F_x, F_x) \\
     &\equiv \frac{\beta}{\mcl{A}}\int^\infty_0\!\mrm{d}t\,\langle F_x(t) F_x(0) \rangle_{\mrm{c}}.
     \label{eqn:GK_friction}
\end{split}
\end{equation}
In the second line, we have used 
$\langle\ldots\rangle_{\rm c}$ to denote
a canonical average in the constrained system.
The lack of correlation in $F_x$ and $P_x$ 
under constrained dynamics, seen in Fig.~\ref{fig1}(b),
means that Eq.~\ref{eqn:uncorrelated} is always satisfied
and there is no negative contribution
to the force autocorrelation function, as seen in Fig.~\ref{fig1}(c).
Therefore, the integral in Eq.~\ref{eqn:GK_friction} always
reaches a well-defined plateau value and does not decay to zero,
as seen in Fig.~\ref{fig1}(d). 
We will show in Sec.~\ref{sec:Interpreting::Effective} that 
$\lambda_{\rm eff}$ measures an effective friction of the entire system and not, in general, the liquid-solid interfacial friction.

Eq.~\ref{eqn:GK_friction} is a core result of the paper, 
presenting a GK relation for liquid--solid
friction where the force autocorrelation function is 
sampled with a zero momentum constraint on the liquid. 
\tcr{We stress that the constraint is proposed as 
a simple way to ensure that the sampled force obeys the orthogonality condition (Eq.~\ref{eqn:uncorrelated}), 
such that the friction coefficient can be obtained from EMD, without explicitly calculating projected forces.
While this constrained ensemble is by no means a
``physical'' model of the system's dynamics, in addition to
numerically verifying its usefulness (see Sec.~\ref{sec:Interpreting}),
we can provide suggestions as to why it might retain many physical
properties relevant for describing liquid--solid friction.}  First,
since the constraint does not change the internal energy of the
system, the hydrodynamic boundary and \tcr{surface roughness\cite{Barrat1999, Falk2010}
remain unchanged between real and
constrained dynamics, as verified by simulations (see Fig.~S3).}
Second, \tcr{when the liquid droplet is constrained to remain at rest,
its center-of-mass motion becomes ``infinitely'' slower compared to
the fast collisions individual liquid particles undergo.  While this}
effectively accelerates the microscopic noise contributing to the
random force, at the microscopic scale, each liquid particle on
average experiences the same potential energy surface with the same
thermal fluctuations, so the overall dissipation in the system should
be largely unaffected provided that liquid droplet is \tcr{sufficiently big.}

\section{Interpreting the friction coefficient}
\label{sec:Interpreting}

\subsection{The effective friction coefficient is not an intrinsic property of the interface}
\label{sec:Interpreting::Effective}

The friction coefficient $\lambda_{\rm eff}$ 
quantifies the dissipation in the system when
there is a net relative motion between the
the liquid and the solid.
When the liquid is pushed out of equilibrium, 
in general, the velocity of the liquid will not be 
constant across the cross-section perpendicular to 
the direction of the flow. Therefore, in addition
to dissipation coming from  collisions of liquid 
particles with the solid at the interface, 
there are also contributions from viscous 
forces arising between adjacent layers of 
the fluid flowing at different velocities
away from the interface. This means that 
$\lambda_{\mrm{eff}}$ will depend on the 
amount of liquid present, and is not an 
intrinsic surface property.
In fact, the effective friction will only be equal to the
intrinsic friction  
$\lambda_{\mrm{eff}}\approx\lambda_{\mrm{intr}}$ either
in the limit of a very thin film of liquid or when there
is perfect slip ($b=\infty$) at the interface 
and the velocity profile becomes plug-like\tcr{; in
such cases, the slip velocity is well approximated by the average fluid velocity, $\overline{v}_x\approx v_\mrm{s}$.}
In this limit, Eq.~\ref{eqn:Markov-droplet} 
becomes 
\begin{equation}
   M \frac{\mrm{d} \overline{v}_x(t)}{\mrm{d}t} = -\lambda_{\mrm{intr}}\mcl{A} v_\mrm{s}(t) + F^{\mrm{R}} (t).
    \label{eqn:limit-slip}
\end{equation}

At this point, it is instructive to point out the difference 
in our derivation compared to that of BB in 
Ref.~\onlinecite{Bocquet2013}. So far, we have worked in the 
solid's frame of reference and treated the liquid's motion as
the coarse-grained variable. We have then applied the 
Langevin equation under the Markovian approximation 
and taken the limit where the velocity profile
in the fluid is a constant (plug flow) to arrive at Eq.~\ref{eqn:limit-slip}. 
Meanwhile, BB started out by considering the solid wall's 
motion as the coarse-grained variable,
analogous to extending Brownian motion to a
planar geometry. In fact, the same expression
for the friction as Eq.~\ref{eqn:GK_friction} 
would result had we chosen to work in the liquid's 
frame of reference and written the Langevin equation as
\begin{equation}
    M_{\mrm{w}} \frac{\mrm{d} U(t)}{\mrm{d}t} = -\lambda_{\mrm{eff}}\mcl{A} U(t) + F^{\mrm{R}} (t),
    \label{eqn:Langevin_wall}
\end{equation}
where $M_{\mrm{w}}$ and $U$ are the mass and velocity of 
the solid wall, respectively. 
\tcr{When the flow profile is plug-like,
$U \to -\overline{v}_x\approx -v_{\mrm{s}}$ and, as discussed above, 
$\lambda_{\mrm{eff}}\approx\lambda_{\mrm{intr}}$.}
In this limit, we recover the Langevin equation as written 
by BB in Ref.~\onlinecite{Bocquet2013}.

To understand why 
the transport coefficient defined by the GK relation in
Eq.~\ref{eqn:GK_friction}
 measures an effective
friction coefficient, not the intrinsic friction
coefficient $\lambda_{\rm intr}$ that enters
Navier's constitutive relation Eq.~\ref{eqn:navier},
it is again helpful to draw analogy to
 Brownian motion: a spherical particle 
moving with constant velocity $\mbf{v}$ 
in the solvent will have a total frictional force
$\mbf{F}=-\xi\mbf{v}$ where the friction
coefficient $\xi$ in this case depends on 
the slippage boundary condition at the sphere's
surface, the sphere's radius and 
the solvent's viscosity.
While this is simply a statement of Stokes' law,
it gives an example of a constitutive
relation relating the total friction with the velocity
of the moving object, not the slip velocity 
at the boundary like in Navier's constitutive relation.

\begin{figure}[t]
  \includegraphics[width=0.8\linewidth]{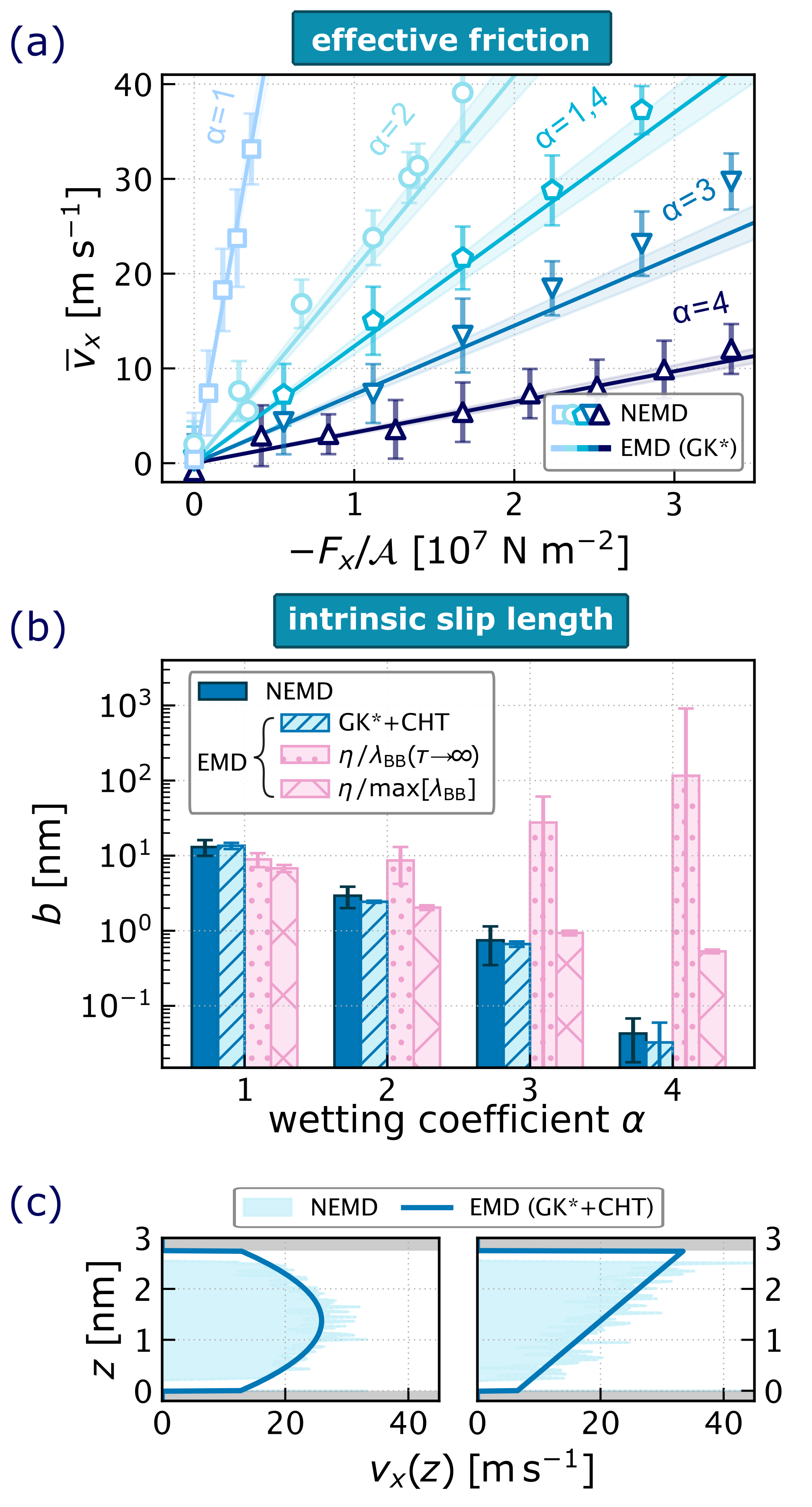}
  \caption{\textbf{Connection between microscopic friction and macroscopic hydrodynamics.} (a) Verification of the constitutive
  relation in Eq.~\ref{eqn:constitutive_relation} between
  the total frictional force and the mean fluid velocity for the effective friction. The data points are from NEMD 
  simulations of Poiseuille flow in 2D channels of water confined
  with surfaces of different wetting coefficients $\alpha$. \tcr{The solid lines have slopes
   $1/\lambda_{\mathrm{eff}}$, where $\lambda_{\mathrm{eff}}$ has been obtained from the GK relation with constrained EMD simulations (Eq.~\ref{eqn:GK_2D}, label GK*).
  (b) The slip lengths obtained from Eq.~\ref{eqn:b_channel}, using $\lambda_{\mathrm{eff}}$ from  EMD simulations (Eq.~\ref{eqn:GK_2D}), labeled GK*+CHT,
  are in excellent agreement with those using
  $\lambda_{\mathrm{eff}}$ from linear fits to NEMD data points in (a).} \tcr{Meanwhile, the performance of the
  BB formula, either by taking the long time limit $\lambda_{\mrm{BB}}(\tau\to\infty)$ or taking the maximum value $\mrm{max}[\lambda_{\mrm{BB}}]$, and interpreting it as an intrinsic friction coefficient (Eq.~\ref{eqn:intrinsic_friction}) gets worse as the surface becomes more attractive.}
  (c) Very good agreement is also obtained for the predicted velocity profile using our approach (GK*+CHT) and reference NEMD data, shown for the case $\alpha = 3$ for both Poiseuille (left) and  Couette (right) flow.
  }
\label{fig2}
\end{figure}

A constitutive relation associated with 
$\lambda_{\rm eff}$ can be obtained using linear 
response theory. To this end, we imagine moving
the liquid droplet at a steady-state velocity of
$\overline{v}_x$ relative to the surface.
The perturbed Hamiltonian is
\begin{equation}
    \mcl{H}_{\mrm{c}}^\prime = \mcl{H}_{\mrm{c}} + \overline{v}_x \hat{\mbf{e}}_x \cdot \sum^N_{i=1} \mbf{p}_{i},
\end{equation}
where $\mcl{H}_{\mrm{c}}$ is the unperturbed Hamiltonian
and $\overline{v}_x \hat{\mbf{e}}_x$ plays the 
role of the external field that couples to the liquid 
momentum  as the conjugate variable. To make a connection with  $\lambda_{\mrm{eff}}$
in Eq.~\ref{eqn:GK_friction},  we consider the response of
$F_x$ under constrained dynamics.
In this case, the momentum conservation constraint
would maintain the liquid's linear momentum at zero $P_x=0$
at equilibrium and at $P_x=M\overline{v}_x$ out-of-equilibrium.
Using linear response theory\cite{EvansMorrissBook} (detailed in the supplementary material),
we show that the response of the tangential frictional
force to a small but finite perturbation  $\overline{v}_x\hat{\mbf{e}}_x$ is
\begin{equation}
   \langle  F_x\rangle^\prime_{\mrm{c}} = -\beta \int^\infty_0\!\mrm{d}t\,\langle F_x(t) F_x(0) \rangle_{\mrm{c}} \overline{v}_x,
   \label{eqn:linear_response}
\end{equation}
where  $\langle\ldots\rangle_{\rm c}^\prime$ and 
$\langle\ldots\rangle_{\rm c}$ denote a constrained 
ensemble average out-of-equilibrium and at equilibrium,
respectively. By combining Eq.~\ref{eqn:linear_response} with the
GK relation in Eq.~\ref{eqn:GK_friction}, we
arrive at a constitutive relation between the
frictional force and the steady-state velocity of the liquid
\begin{equation}
    F_x = - \lambda_{\mrm{eff}}\mcl{A}\overline{v}_x.
    \label{eqn:constitutive_relation}
\end{equation}
Here, we have dropped the canonical average on the force
when writing the constitutive relation.
This expression is in fact analogous to Stoke's law,
and $\lambda_{\mrm{eff}}$ therefore depends not only on the
slippage boundary conditions and the viscosity of the liquid,
but also the size and shape of the liquid. 

\tcr{Applying the framework that we have derived to confined fluids is
straightforward; the total frictional force on the liquid ($F_x$) now
contains contributions from all solid surfaces in contact with the
fluid, while $v_x$ is simply the average flow velocity of the fluid.}
To verify Eq.~\ref{eqn:constitutive_relation}, we considered systems
with water confined in four symmetric channels made up of solid
substrates with attractive strengths of $\varepsilon_{\rm
wf}=\alpha\varepsilon_{0}$ where the ``wetting coefficients" are
$\alpha=1,2,3,4$ and $\varepsilon_{0}=1.57\,\mrm{kJ\,mol^{-1}}$.  We
also consider one asymmetric channel with $\alpha=1$ for the top wall
and $\alpha=4$ for the bottom wall.  Various channel heights between the first atomic
planes of the solids are considered, $H/\mrm{nm}\approx 1.4, 2.7, 5.2$.  Details on the systems and molecular models can
be found in the supplementary material.  Using constrained EMD
simulations, we computed $\lambda_{\mrm{eff}}$ from the GK relation in
Eq.~\ref{eqn:GK_friction}.
\tcr{For reference NEMD simulations, we performed
simulations of Poiseuille flow by applying a body force on each oxygen atom to 
mimic the effect of a pressure gradient. 
In the regime of low driving force, 
the linear response predicted by $\lambda_{\mrm{eff}}$
obtained from constrained EMD simulations
(labeled GK*)
is in excellent agreement with 
$\overline{v}_{x}$ 
measured from NEMD simulations,}
shown for channels of different  attractive strengths in
Fig.~2(a), and for channels of different 
heights in the supplementary material.  
In the supplementary material \tcr{(Fig.~S3)}, we also confirmed 
the equivalence between the equilibrium and
out-of-equilibrium force \tcr{auto}correlation, as expected 
from the fluctuation--dissipation theorem  in the linear
response regime.

\subsection{Connection to the hydrodynamic slip boundary}
\label{sec:Interpreting::Connection}

While the GK relation provides a microscopic
expression for the effective friction, more often than not,
one is interested in the intrinsic friction or 
the slip length of a particular interface. 
Since the constitutive relation of the effective friction has
been established, it is relatively straightforward to 
obtain a relationship between the effective friction and 
the slip length from macroscopic hydrodynamics. 
Specifically, we can solve the Navier--Stokes 
equation\cite{LandauLifshitzBook,birdBook} subject to partial
slip boundary conditions like Eq.~\ref{eq:boundary} at the solid wall
to obtain the Poiseuille flow profile of the fluid.
From the solution, we then obtain 
the fluid velocity as well as its frictional  force per unit area,
the ratio of which gives an expression for the effective
friction coefficient. Here, we will simply state the results for the common cases of 2D and 1D confinement, with details presented in the supplementary material.

For 2D confinement, the CHT result
for the effective friction of a fluid of shear viscosity
$\eta$ confined in a channel of height $H$,
made up of two separate interfaces is 
\begin{equation}
    \lambda^{\mrm{2D}}_{\mrm{eff}}(H) = \frac{12(H + b_1 + b_2)\eta}{H^2 + 4 H (b_1 + b_2) + 12 b_1 b_2},
    \label{eqn:2D_friction}
\end{equation}
where $b_1$ and $b_2$ are the slip lengths of each interface.
To obtain a closed expression for the slip length
of a single interface, 
we can consider the case when the channel is symmetric, i.e.
$b_1 = b_2 = b$, and rearrange Eq.~\ref{eqn:2D_friction},
\begin{equation}
\begin{split}
    b  = & \left(\frac{\eta}{\lambda^{\mrm{2D}}_{\mrm{eff}}(H)} - \frac{H}{3}\right) \\
    & + \left[ \left(\frac{\eta}{\lambda^{\mrm{2D}}_{\mrm{eff}}(H)} - \frac{H}{3}\right)^2 +H\left(\frac{\eta}{\lambda^{\mrm{2D}}_{\mrm{eff}}(H)} - \frac{H}{12}\right)  \right]^{1/2}.
    \label{eqn:b_channel}
\end{split}
\end{equation}
The effective friction in a 2D channel 
can be computed 
with the GK expression 
\begin{equation}
   \lambda^{\mrm{2D}}_{\mrm{eff}} =   \frac{\beta}{2\mcl{A}}\int^\infty_0\!\mrm{d}t\,\langle \mbf{F}_{\parallel}(t) \cdot \mbf{F}_{\parallel}(0) \rangle_{\mrm{c}},
   \label{eqn:GK_2D}
\end{equation}
from an EMD simulation with a constraint keeping
 the liquid's linear momentum zero 
 in the in-plane $(x,y)$ directions, 
where $\mbf{F}_{\parallel}=(F_x,F_y)$ denotes 
 the in-plane lateral force on the liquid.

For 1D confinement, we consider Poiseuille
flow of a fluid through a cylindrical tube
of radius $R$. In an analogous manner, 
the effective friction coefficient can be 
obtained from classical hydrodynamics as
\begin{equation}
    \lambda^{\mrm{1D}}_{\mrm{eff}}(R) = \frac{4\eta}{R + 4b_R}.
        \label{eqn:1D_friction}
\end{equation}
where the slip length $b_R$ is in 
general dependent on the curvature and therefore the
radius of the tube.  Rearranging for $b_R$, we obtain
\begin{equation}
    b_R = \frac{\eta}{\lambda^{\mrm{1D}}_{\mrm{eff}}(R)} - \frac{R}{4},
    \label{eqn:b_tube}
\end{equation}
where $\lambda^{\mrm{1D}}_{\mrm{eff}}$ can be obtained with
\begin{equation}
   \lambda^{\mrm{1D}}_{\mrm{eff}} =   \frac{\beta}{\mcl{A}}\int^\infty_0\!\mrm{d}t\,\langle F_{z}(t) F_{z}(0) \rangle_{\mrm{c}},
   \label{eqn:GK_1D}
\end{equation}
from an EMD simulation with a constraint keeping the liquid's 
momentum zero along the axial direction ($z$) of the tube,
where $F_z$ denotes the force on the liquid in the axial direction.

The pair of Eqs.~\ref{eqn:b_channel}/\ref{eqn:GK_2D} and \ref{eqn:b_tube}/\ref{eqn:GK_1D} provides 
a way to obtain hydrodynamic slippage directly from
fluctuating forces in \tcr{microscopic EMD simulations. 
In Fig.~\ref{fig2}(b), we validate that the slip lengths $b$ 
obtained by this method (labeled GK*+CHT) are in excellent agreement with results from reference NEMD 
simulations for surfaces of different wettability. (To obtain $b$ from NEMD, we have measured $\lambda_{\rm eff}$ and 
used the result in the CHT expression, rather than attempt to fit the flow profiles directly.) In contrast, using 
the BB expression in Eq.~\ref{eqn:BB} combined with  
Eq.~\ref{eqn:intrinsic_friction}, as is typically done in 
the literature, significantly overpredicts
the slip length as the wettability increases.}


\tcr{While sources of uncertainty in the effective friction measured with the GK
relation arise from statistical error, the value of slip length is
also impacted by the choice of hydrodynamic boundary position $z_0$
(which enters Eqs.~\ref{eqn:b_channel} and \ref{eqn:b_tube} implicitly
through $H$ and $R$).
While approaches exist for determining $z_0$, see, e.g.,
Refs.~\onlinecite{Herrero2019} and \onlinecite{Camargo2019}, here we make the pragmatic choice to
place $z_0$ at the first plane of solid atoms in contact with the
liquid. With this choice,} the solution for the velocity profile from
the Navier--Stokes equation spans the total height of the channel or
the diameter of the cylindrical tube.  However, microscopically, there
is potentially an offset in the location of the hydrodynamic boundary
due to the excluded volume at each liquid--solid interface.
\tcr{In the supplementary material, 
we assess the sensitivity of our results to such an offset. We find
that, when the fluid's friction is dominated by viscous dissipation
(i.e., in tubes or channels that are large compared to the slip
length), results are robust to reasonable choices of $z_0$. As $H$ or
$R$ tends to zero, obtained values of $b$ become increasingly sensitive
to the choice of $z_0$; generally speaking, in cases where
$H>1\,\mrm{nm}$, placing the hydrodynamic boundary at the first
atomic plane of the solid is a reasonable approximation.}

It is important to note  that the appropriate type 
of flow used to give the mapping between CHT and the
effective friction is Poiseuille flow, where the fluid
has been driven by a pressure gradient or a body force.
This is because under confinement, when the liquid is 
in contact with two separate solid substrates, the
solid's frame of reference is only well-defined when
both substrates have the same velocity as each other.
\tcr{Note that the mapping to Poiseuille flow enables calculation 
of friction in cylindrical geometries.}
Meanwhile, in the case of Couette flow, the two solid 
substrates are moving at different velocities 
in order to shear the fluid. 
However,
as the slip length $b$ is an intrinsic property, once 
its value is known, 
predictions of the velocity profile can be obtained from CHT.
For example, we show the predictions for 
the symmetric $\alpha = 3$ channel, both for Couette
flow and Poiseuille flow in Fig.~\ref{fig2}(c), 
and for other cases extensively in the supplementary material. In all cases,  
we find very good agreement with reference NEMD simulations.

\section{Application: Water friction under confinement}

\begin{figure*}[t]
  \includegraphics[width=0.95\linewidth]{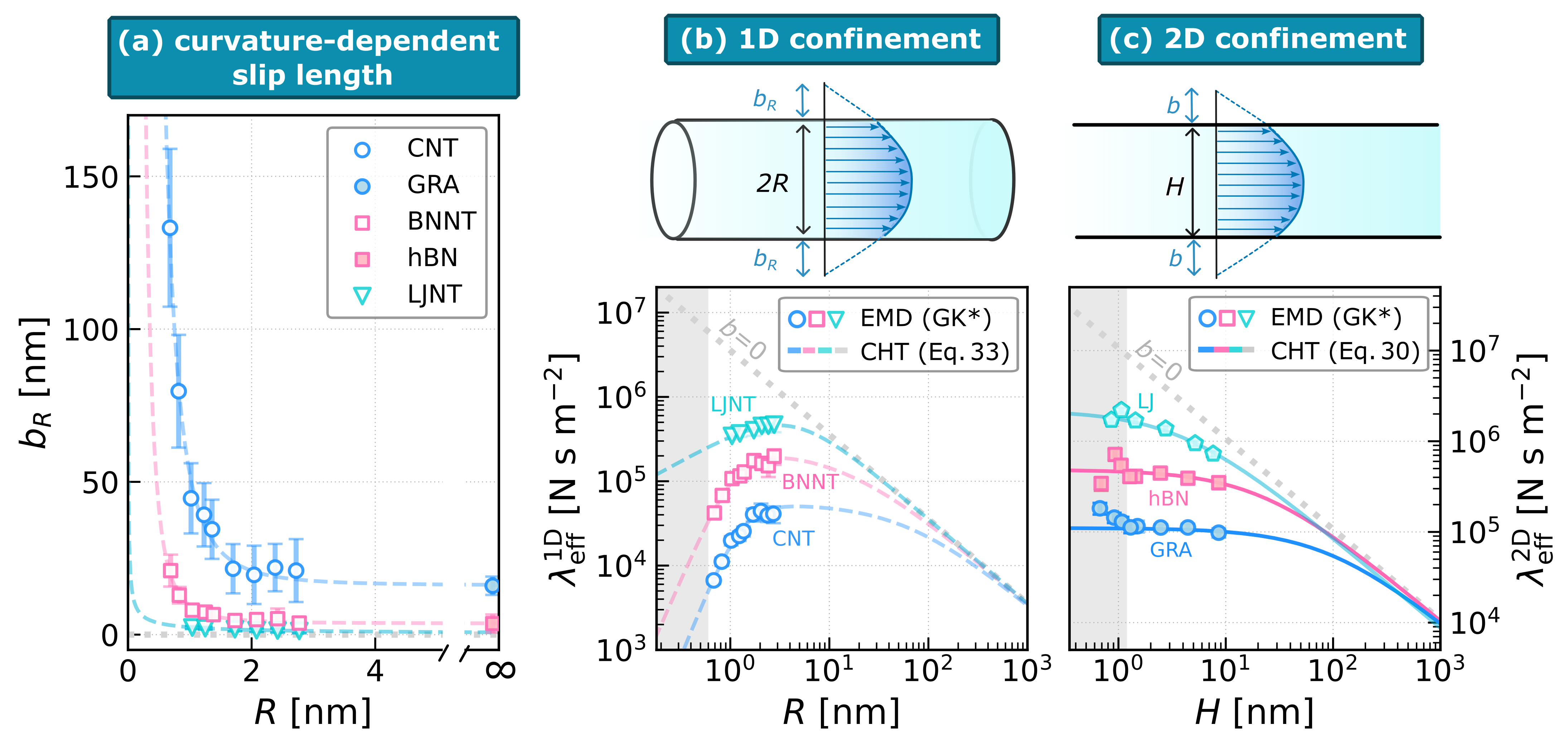}
  \caption{\textbf{Water friction under confinement.} (a) The curvature dependence of the slip length $b$ for water on different surfaces including carbon nanotubes (CNTs), boron nitride nanotubes (BNNTs), nanotubes made of  Lennard--Jones particles (LJNTs), 
  a flat graphene sheet (GRA) and a flat hexagonal boron nitride sheet (hBN) obtained from simulations.  The effective friction
  under 1D and 2D confinement
  is shown in (b) and (c) as a function of the tube radius $R$ and channel height $H$. 
  \tcr{The data points are obtained
  by the GK relation using constrained EMD simulations.
  For 2D confinement, the solid lines are predictions from CHT (Eq.~\ref{eqn:2D_friction}) parameterized
  on the slip length of the largest channel.
  For 1D confinement, the dashed lines are from Eq.~\ref{eqn:1D_friction} using $b_R(R)$ fitted to the simulation data.}
  The \tcr{dotted} lines show the results using
  no-slip boundary conditions while solid lines account for finite
  slippage at the interface. The shaded regions indicate where
  CHT begins to break down.}
\label{fig3}
\end{figure*}

Having a connection between a macroscopic and 
microscopic description of fluid flow, we can
revisit the long-standing question on the domain of 
applicability of the continuum hydrodynamic equations.
Such a question is two-fold, as it concerns both the
validity of the boundary conditions applied when
solving the Navier--Stokes equation and the validity of 
continuum hydrodynamics itself.

Historically, the no-slip boundary condition $b=0$
can successfully describe much of everyday phenomena 
involving fluid dynamics \cite{feynman1963feynman,goldstein1950modern}.
For water flowing through macroscopic tubes and channels,
it provides a very good approximation, regardless of the 
material of the solid wall. At these larger scales,
the effective friction increases monotonically as the
tube becomes smaller $\lambda^{\mathrm{1D}}_{\rm eff}\sim 4\eta/R$ 
or as the channel height decreases 
$\lambda^{\mrm{2D}}_{\rm eff}\sim12\eta/H$.
\tcr{It has also long been established that as the
length scale of confinement keeps shrinking 
in size, finite slippage at the interface can no longer
be ignored  \cite{Zhu2002,Thompson1997} as a
larger proportion of the confined
fluid can feel the interface.} Thus, the first question we
will address is \textit{``At which confinement length scale 
do surface effects begin to manifest as a
deviation from the no-slip boundary approximation?''} 
The answer will depend on various factors, including 
the material and the curvature of confinement.

To answer this question, we will consider 1D confinement
in single-walled carbon and boron nitride nanotubes
of different radii and 2D confinement in channels 
of different heights formed by two graphene or 
boron nitride sheets. For the interatomic interaction, 
we employ a machine-learned potential where the electronic 
structure is accurately trained at the level of density 
functional theory \cite{Thiemann2022};
\tcr{the walls are flexible in these simulations.}
While we consider these low-dimensional materials for
their relevance to applications in nanofluidics, we 
stress the generality of the results obtained to 
confinement by other solid surfaces of other liquids. 
While the microscopic mechanisms giving rise to friction
at the interface are very diverse 
(corrugation \cite{Thiemann2022, Falk2010}, 
phonon coupling \cite{Ma2016}, 
quantum friction \cite{Kavokine2022, Bui2023}, 
defects \cite{Joly2016, Seal2021, Li2024}, etc.) 
and  we expect them to depend sensitively on
the molecular details of the interface, the macroscopic 
hydrodynamic behavior would be principally governed by 
the slip length. Therefore, in addition to these 
realistic surfaces, we also consider water
confined in Lennard--Jones nanotubes and 
between walls of different wettability.

For each surface, we first obtain
the slip length from EMD simulations using
the pair of Eqs.~\ref{eqn:b_channel}/\ref{eqn:GK_2D} and \ref{eqn:b_tube}/\ref{eqn:GK_1D}. 
Between the different wall materials,
the graphene surface has the highest slip length
($b=16\,$nm), followed by boron nitride ($b=3.6\,$nm) 
and the Lennard--Jones surface of high 
wettability where there is essentially no slippage.
In the case of 1D confinement, in addition to surface
effects due to different confining materials, curvature
effects also come into play since curving up a surface
can change its free energy surface significantly.
We show the curvature-dependent slip length $b_R$ for three 
different surfaces in Fig.~3(a). In all cases, curving 
up the surface to form nanotubes leads to a decrease 
in the corrugation of the surface, reflected by an 
increase in $b_R$ as $R$ decreases. This enhancement
is most prominent for tubes of $R\lesssim 5\,$nm with
the slippery carbon surface. Our values are broadly 
in agreement with previous simulations and 
experiments of water on graphene and boron nitride surfaces
\cite{Falk2010,Tocci2014,Secchi2016,Keerthi2021,Thiemann2022,Bui2023}.

To assess the overall effect of changing the radius on 
water flow in 1D confinement, using $b_R(R)$ obtained from
a non-linear fit of the simulation data in 
Eq.~\ref{eqn:2D_friction}, we show in Fig.~\ref{fig3}(b)
the effective friction $\lambda^{\mrm{1D}}_{\mrm{eff}}$ 
as a function of $R$. There are two competing effects
dictating the behavior of $\lambda^{\mrm{1D}}_{\mrm{eff}}$.
For tubes with diameters on the micrometer scale and beyond,
$\lambda^{\mrm{1D}}_{\mrm{eff}}$ increases as $R$ 
decreases for all surfaces, asymptotic to the no-slip prediction.
As $R$ reaches the nanometric regime, since it
is now comparable to $b_R$, the effective friction
in the fluid is no longer agnostic to the amount
of slippage taking place at the wall. As a result, the
enhancement in $b_R$ due to curvature effects 
starts to dominate, decreasing 
$\lambda^{\mrm{1D}}_{\mrm{eff}}$ as $R$ decreases.
Therefore, a crossover in the dependence of
$\lambda^{\mathrm{1D}}_{\rm eff}$
on $R$ can be observed at $R\sim10\,$nm and 
is most significant for surfaces of the highest slippage.
We note that experiments \cite{Secchi2016} with multi-walled 
carbon nanotubes have reported a radius dependence of 
slip length up to $50\,$nm, so the length scale of 
the crossover could be as large as tens of nm.
For 2D confinement, since there is no curvature
effect, any surface effect would be solely 
due to interactions between the solid and the fluid. 
In this case, instead of a crossover, 
a deviation to the no-slip prediction
is seen for the asymptotic behavior of
$\lambda^{\mrm{2D}}_{\mrm{eff}}$ as 
the channel height $H$ decreases, 
as shown in Fig.~3(c) for water confined in 
symmetric channels of graphene, 
hexagonal boron nitride and the Lennard--Jones 
solid substrates.
Similarly, the length scale of this deviation is in 
the tens of nm and is most significant for
surfaces of the highest slippage.

Hydrodynamics relies on the key assumption 
that the fluid behaves as a continuum in the hydrodynamic 
regime, where the local properties of the fluid
vary slowly on microscopic time and length scales 
\cite{HansenMcDonaldBook}. As the degree of confinement 
increases, this approximation is expected to break down 
as the separation in scale between the confinement 
length and the molecular length becomes less clear.
This naturally brings us to the second question:
\textit{``Once finite slippage is accounted for at the 
interface, what is the confinement length scale at which
it is still possible to describe flow with continuum 
hydrodynamics?}''

\tcr{Since the GK relation we have derived is independent of CHT, we can
parameterize CHT with the slip length obtained
from a particular geometry, and then assess how well this parameterized CHT performs as we decrease the confinement length.
To this end, we compare the effective friction predicted by CHT (Eq.~\ref{eqn:2D_friction}) 
with that computed from the GK relation for various channel heights down to
the sub-nanometer regime, shown as data points in Fig.~\ref{fig3}(c).}
We find that in all confining materials considered,
CHT predictions hold remarkably well down to $H\sim1\,$nm, where the 
channels can accommodate only three layers of water. 
The deviation between CHT and simulations
is only significant for smaller channels  with 
a  bilayer and monolayer of water, where we find 
that $\lambda^{\mathrm{2D}}_{\rm eff}$ from simulations 
is extremely sensitive to changes in the density
and interfacial structure of the fluid.
The breakdown of CHT below $1\,$nm is perhaps not 
surprising since with fewer than three layers of water,
one can no longer sensibly define a continuum bulk region.
In fact, the length scale of $1\,$nm has been suggested 
both experimentally \cite{Raviv2002,Li2007} and theoretically 
\cite{Leng2005, Bocquet2010} as the limit for the validity
of the notion of a bulk shear viscosity.
Nevertheless, the robustness of a continuum theory to
describe water flow down to such a small scale is extremely 
satisfying. It is also reminiscent of the validity 
of other macroscopic theories, e.g., continuum 
mechanics \cite{Mo2009}, the Kelvin equation \cite{Yang2020}
or dielectric continuum theory\cite{Cox2022, Loche2020,Cox2018, Schlaich2016}, 
down to the nanoscale.

\section{Conclusion}

In this work, through the use of the projection operator
formalism and linear response theory, we have derived a
GK relation for the liquid--solid friction 
by considering the stochastic motion of a liquid droplet
on a solid surface. The force autocorrelation function in the
expression is sampled using constrained dynamics 
in which the momentum of the liquid is conserved.
Not only does such an expression help resolve the
long-standing plateau problem associated with previous 
studies of friction using molecular simulations, but it also
sheds light on the physical meaning of the obtained friction 
coefficient: $\lambda_{\rm eff}$ measures 
an effective friction and not the intrinsic interfacial friction.
Our expression can be applied generally to
finite systems with surfaces of very low or
very high friction. In contrast, the widely used 
expression by Bocquet and Barrat was derived originally
for a semi-infinite
system\cite{Bocquet1997, Bocquet2013}. When applied to finite systems, in the context of our work, it amounts to a reasonable approximation only when 
the slip length is much greater than the confining length scale.

By linking the microscopic expression for friction to macroscopic
hydrodynamics, we show how the hydrodynamic slip length
can be obtained faithfully, both for 1D and 2D fluid flow.
Going forwards, we believe that such a connection is 
important to understand the flow of liquids
across length scales on more complex surfaces 
where slippage is determined by the interplay of 
the material's electronic properties,
wettability, surface charge and the
presence of defects. This work also lays the foundation 
for understanding  interfacial effects in 
electrokinetic phenomena
and other transport properties under confinement 
\cite{Joly2006,Simonnin2017}.
\tcr{We believe our work can provide insights for
the development of microscopic descriptions
of hydrodynamics based on dynamical density functional 
theory\cite{Camargo2018, Camargo2019, Marconi2000, Espanol2009},
the non-equilibrium extension
of equilibrium classical density functional theory\cite{Evans1979, HansenMcDonaldBook, Bui2024},
in which the fluid flow is described with mass and momentum 
density fields.}

Finally, as an application of our approach, 
we assess the ability of continuum hydrodynamics 
in describing water flow at the nanoscale. 
\tcr{While for macroscopic confining geometries assuming
no-slip boundary conditions is a reasonable approximation, the impact
of finite slippage cannot be ignored for confinement length scales
below 10--100\,nm. Nevertheless, our results show that, provided the
finite slip length is taken into account, continuum hydrodynamics
remains remarkably robust down to $\sim$1\,nm confinement.}



\section*{Data Availability}

\normalsize

The data that support the findings of this study are openly available at the University of Cambridge Data Repository at  \url{https://doi.org/10.17863/CAM.111725}.
Scripts for computing the friction coefficient are available at
\url{https://github.com/annatbui/friction-GK}.

\section*{Supplementary Material}

Supplementary material includes simulations details and additional supporting results.

\section*{Acknowledgements}

We thank Mathieu Salanne, Laurent Joly and
Daan Frenkel for insightful discussions.
Via membership of the UK's HEC Materials
Chemistry Consortium funded by EPSRC
(EP/X035859), this work used the ARCHER2 UK
National Supercomputing Service.
A.T.B. acknowledges funding from the Oppenheimer Fund
and Peterhouse College, University of Cambridge.
S.J.C. is a Royal Society University Research
Fellow (Grant No. URF\textbackslash
R1\textbackslash 211144).

\bibliography{references}

\end{document}


\title{Supplementary material: Revisiting the Green--Kubo relation for friction in nanofluidics}

\author{Anna T. Bui}
\affiliation{Yusuf Hamied Department of Chemistry, University of
  Cambridge, Lensfield Road, Cambridge, CB2 1EW, United Kingdom}

\author{Stephen J. Cox}
\email{stephen.j.cox@durham.ac.uk}
\affiliation{Yusuf Hamied Department of Chemistry, University of
  Cambridge, Lensfield Road, Cambridge, CB2 1EW, United Kingdom}
\affiliation{Department of Chemistry, Durham University, South Road, Durham, DH1 3LE, United Kingdom}

\date{\today}

\maketitle

\tableofcontents
\newpage

\section{Derivation details}

\subsection{Green--Kubo friction}


As in the main paper, we consider the problem of the a liquid 
droplet's stochastic motion on a solid surface. 
The hamiltonian for the system of $N$ liquid particles 
whose positions and momenta are $\{\mbf{r}_i,\mbf{p}_i\}$ and
masses are $m_i$ is
\begin{equation}
    \mcl{H} = \sum^{N}_{i=1} \frac{\mbf{p}_i^2}{2m_i} + V(\mbf{r}^N) + V_{\mrm{ext}}(\mbf{r}^N),
\end{equation}
where $V$ is the interparticle potential energy and $V_{\mrm{ext}}$ is the potential
energy from interaction of the particles with the solid wall.
We are interested in the dynamical variable $P_x(t)$
which is the momentum of the liquid droplet
in the $x$ direction. Ordinarily, the time 
evolution of $P_x$ is governed by Liouville equation
\begin{equation}
    \frac{\mrm{d}P_x(t)}{\mrm{d}t}= i\mcl{L}P_x(t), \quad\quad P_x(t)=\mrm{e}^{i\mcl{L}t}P_x(0), \quad\quad\mrm{[real\,dynamics]}
\end{equation}
where the Liouville operator $i\mcl{L}$ specifies the ``real dynamics'' that the system undergoes when no constraints are applied.
Explicitly, $i\mcl{L}$ is given with the Poisson bracket
\begin{equation}
    i\mcl{L} \equiv \{ \mcl{H}, \boldsymbol{\cdot} \} = \sum^N_{i=1}\left( \frac{\mbf{p}_i}{m_i}\cdot \frac{\partial }{\partial \mbf{r}_i} + \mbf{f}_i\cdot \frac{\partial }{\partial \mbf{p}_i}\right),
    \label{eqn:unconstrained-Liouville}
\end{equation}
where $\mbf{f}_i=-\partial V / \partial \mbf{r}_i  -  \partial V_{\mrm{ext}} / \partial \mbf{r}_i $  
is the force on each particle.
For any variable $X$, to consider the projected part of $X$ on $P_x$, we denote the projection operator $\mcl{P}$ as
\begin{equation}
    \mcl{P} X(t) = (P_x,P_x)^{-1} (X(t),P_x) P_x,
\end{equation}
such that
\begin{equation}
    (\mcl{P} X(t), P_x) = (X(t),P_x) \equiv \langle X(t)P_x(0) \rangle,
\end{equation}
where $(\cdots,\cdots)$ denotes a scalar product
and $\langle \cdots \rangle$ denotes an ensemble average
at equilibrium.
The complementary operator $\mcl{Q}=\mathds{1}-\mcl{P}$ projects onto the subspace orthogonal to $P_x$, such that
\begin{equation}
    ( \mcl{Q}X(t), P_x ) \equiv \langle \mcl{Q}X(t) P_x(0) \rangle = 0.
\end{equation}

As we showed in the main article, the effective friction coefficient of the droplet is given as 
\begin{equation}
\begin{split}
    \lambda_{\mrm{eff}} & = \frac{\beta}{\mcl{A}} \int^{\infty}_0\!\mrm{d}t\, ( \mrm{e}^{i\mcl{QL}t}F_x, F_x ) \\
    & \neq \frac{\beta}{\mcl{A}} \int^{\infty}_0\!\mrm{d}t\, \langle F_x(t) F_x(0) \rangle.
\end{split}
\end{equation}
In general, the projected dynamics is unknown. 
Instead we can make use of the result in Ref.~\onlinecite{Hijon2010} which shows that
the projected dynamics can be made explicit by 
introducing appropriate constraints. The effect
of the constraints is to enforce the appropriate
separation of timescales.

To be precise, we seek a constrained dynamics with a 
Liouville operator that satisfies
\begin{equation}
    i\mcl{P}\mcl{L}_{\mrm{c}} = 0,\quad\quad i\mcl{PQ(\mcl{L}-\mcl{L}_{\mrm{c}})\mcl{P}} = 0.
    \label{eq:conditions}
\end{equation}
The modified dynamics can be formally introduced with
the following Liouville operator
\begin{equation}
    i\mcl{L}_\varepsilon =    i\mcl{P}(\mcl{L}-\mcl{L}_{\mrm{c}}) + \frac{1}{\varepsilon}i\mcl{Q}(\mcl{L}-\mcl{L}_{\mrm{c}})+\frac{1}{\varepsilon^2}i\mcl{L}_{\mrm{c}},
\end{equation}
where $0<\varepsilon<1$. In the limit $\varepsilon\to 1$, $i\mcl{L}_{\varepsilon}\to i\mcl{L}$
so the original dynamics is recovered. In the limit 
$\varepsilon\to 0$, the Markovian approximation is
enforced, which can be seen by considering the generalized 
Langevin equation
\begin{equation}
    M \frac{\mrm{d}\overline{v}_x(t)}{\mrm{d}t} = -\int^t_0\!\mrm{d}t^\prime\, K_{\varepsilon}(t^\prime) \overline{v}_x(t-t^\prime) + F^{\mrm{R}}(t),
    \label{eqn:beforechange}
\end{equation}
where $K_{\varepsilon}$ is the memory function and $F^{\mrm{R}}$ is random force.
With a change in 
integration variable  $t^\prime = \varepsilon^2 \tau$,  Eq.~\ref{eqn:beforechange} becomes
\begin{equation}
    M \frac{\mrm{d}\overline{v}_x(t)}{\mrm{d}t} = -\int^{t/\varepsilon^2}_0\!\mrm{d}\tau\, \varepsilon^2K_{\epsilon}(\varepsilon^2 \tau) \overline{v}_x(t-\varepsilon^2 \tau) + F^{\mrm{R}}(t).
    \label{eqn:afterchange}
\end{equation}
The memory kernel in the modified dynamics reads
\begin{equation}
  K_{\varepsilon}(\varepsilon^2\tau) = \beta( F^{\mrm{R}}(\varepsilon^2\tau),  F^{\mrm{R}})=\beta( \mrm{e}^{i\mcl{QL}_{\varepsilon}\varepsilon^2\tau}i\mcl{QL}_{\varepsilon} P_x, i\mcl{QL}_{\varepsilon} P_x). 
\end{equation}
The main result that we use from Ref.~\onlinecite{Hijon2010} is that 
\begin{equation}
  \lim_{\varepsilon\to0}\varepsilon^2K(\varepsilon^2 \tau) = \beta ( \mrm{e}^{i\mcl{L}_{\mrm{c}} \tau} i\mcl{L }P_x, i\mcl{L} P_x) \equiv \beta \langle F_x(\tau) F_x(0)\rangle_{\mrm{c}},
\end{equation}
where $\langle\cdots\rangle_{\mrm{c}}$ denotes a canonical average in the 
constrained system. 
As the Markovian limit has been imposed by the constraints, we now have
a simple Langevin equation
\begin{equation}
    M \frac{\mrm{d}\overline{v}_x(t)}{\mrm{d}t} = - \lambda_{\rm eff}\mcl{A}\overline{v}_x(t) + F^{\mrm{R}}(t),
\end{equation}
where the effective friction coefficient is given as 
\begin{equation}
    \lambda_{\rm eff} = \frac{\beta}{\mcl{A}}\int^\infty_0\!\mrm{d}\tau\,\langle  F_x(\tau) F_x(0)  \rangle_{\mrm{c}}.
    \label{eq:GK_friction}
\end{equation}
In the next section, we will explicitly show what
the constrained Liouville operator $i\mcl{L}_{\mrm{c}}$ is.

\subsection{Constrained Liouvillian}

To satisfy Eq.~\ref{eq:conditions}, the constrained Liouville operator
should enforce the constraint of conserving $P_x$
\begin{equation}
    \frac{\mrm{d}P_x(t)}{\mrm{d}t}= i\mcl{L}_{\mrm{c}}P_x(t) = 0, \quad\quad\mrm{[constrained\,dynamics]}
\end{equation}
and at equilibrium, the liquid droplet on average has zero linear momentum.
In other words, the liquid droplet's center of mass position
is constrained to its initial positions.  The constraint functions
of phase space are therefore
\begin{equation}
\begin{split}
   R_x &= \frac{1}{M} \sum^N_{i=1} m_i \dot{\mbf{r}}_i \cdot \hat{\mbf{e}}_x = \mrm{constant}, \\
P_x &= \sum^N_{i=1} m_i \dot{\mbf{r}}_{i} \cdot \hat{\mbf{e}}_x =0,
\end{split}
\end{equation}
where $R_x$ denotes the position of the center of mass of the droplet.
To determine how the Hamiltonian is modified by the constrained, we employ
Lagrange's method of undetermined multipliers. The constrained Hamiltonian
is given as
\begin{equation}
    \mcl{H}_{\mrm{c}} = \mcl{H} - \mu \hat{\mbf{e}}_x \cdot \sum^{N}_{i=1} m_i \dot{\mbf{r}}_i + \gamma \hat{\mbf{e}}_x\cdot  \sum^{N}_{i=1} m_i \mbf{r}_i,
\end{equation}
where $\mu$ and $\gamma$ are the Lagrange multipliers. The corresponding equations of motion are 
\begin{equation}
\begin{split}
    \dot{\mbf{r}}_i & = \frac{\partial \mbf{H}_{\mrm{c}}}{\partial \mbf{p}_i} =  \frac{\mbf{p}_i}{m_i} - \mu\mbf{e}_x, \\ 
    \dot{\mbf{p}}_i & = -\frac{\partial \mbf{H}_{\mrm{c}}}{\partial \mbf{r}_i} = \mbf{f}_i - \gamma m_i \mbf{e}_x.
    \label{eqn:equations_of_motion}
\end{split}
\end{equation}
To determine $\mu$ and $\gamma$, we consider the time derivative of the constraint functions
\begin{equation}
\begin{split}
    \dot{R}_x &= 0 \quad \Rightarrow \quad  \sum^N_{i=1} m_i \dot{\mbf{r}}_i \cdot \hat{\mbf{e}}_x = 0, \\
     \dot{P}_x &= 0  \quad \Rightarrow \quad  \sum^N_{i=1} \dot{\mbf{p}}_i \cdot \hat{\mbf{e}}_x = 0.
     \label{eqn:constraints_derivative}
\end{split}
\end{equation}
Upon insertion of Eq.~\ref{eqn:equations_of_motion} into Eq.~\ref{eqn:constraints_derivative}, we find
\begin{equation}
    \mu = \frac{P_x}{M}, \quad\quad \gamma = \frac{F_x}{M}.
\end{equation}
With $\mu$ and  $\gamma$ determined, the constrained  Lioville operator can
be written as
\begin{equation}
    i\mcl{L}_{\mrm{c}} \equiv \{ \mcl{H}_{\mrm{c}}, \boldsymbol{\cdot} \}= \sum^N_{i=1}\left[ \left(\frac{\mbf{p}_i}{m_i} - \mu\hat{\mbf{e}}_x \right)\cdot \frac{\partial }{\partial \mbf{r}_i} + \left(\mbf{f}_i - \gamma m_i\hat{\mbf{e}}_x\right)\cdot \frac{\partial }{\partial \mbf{p}_i}\right].
    \label{eqn:constrained-Liouville}
\end{equation}

\subsection{Linear response } 

In this section, we report a more detailed derivation of the constitutive relation in 
Eq.~\tcr{29} of the main text which frictional force $F_x$ and the
average velocity of the fluid $\overline{v}_x$. To do so, we will follow
Evans and Morriss' treatment of linear response theory\cite{EvansMorrissBook} and 
consider a perturbed Hamiltonian $\mcl{H}^\prime$
of the form
\begin{equation}
    \mcl{H}_{\mrm{c}}^\prime(\Gamma) = \mcl{H}_{\mrm{c}}(\Gamma) + \overline{v}_x \hat{\mbf{e}}_x \cdot \sum^N_{i=1} \mbf{p}_{i},
\end{equation}
where $\mcl{H}_{\mrm{c}}$ is the unperturbed Hamiltonian for the liquid
droplet constrained at rest $P_x(t)=0$ and $\Gamma=(\mbf{r}^N,\mbf{p}^N)$ denotes the phase space of the liquid. The perturbation takes the droplet to a finite steady state velocity such that the momentum is now constrained at $P_x(t)=M\overline{v}_x$ so $\overline{v}_x \hat{\mbf{e}}_x$ plays the role of the external field and the sum over $N$ liquid particle momenta $\mbf{p}_{i}$ is the corresponding conjugate variable.

Explicitly, the constrained dynamics is propagated by the
constrained perturbed Liouville operator
\begin{equation}
     i\mcl{L}^\prime_{\mrm{c}} \equiv \{ \mcl{H}^\prime_{\mrm{c}}, \boldsymbol{\cdot} \} =  i\mcl{L}_{\mrm{c}} + i\Delta\mcl{L}^\prime_{\mrm{c}},
\end{equation}
where the field-free part $i\mcl{L}_{\mrm{c}}$ is given as before in Eq.~\ref{eqn:constrained-Liouville}.
The field-dependent part is
\begin{equation}
    i\Delta\mcl{L}^\prime_{\mrm{c}}   = \sum^N_{i=1}\left[ \overline{v}_x\hat{\mbf{e}}_x \cdot \frac{\partial }{\partial \mbf{r}_i} \right],
\end{equation} 
The perturbed equilibrium distribution function $f^\prime(\Gamma,t)$ 
of the statistical ensemble
sampled can be
obtained by solving the following Liouville equation 
\begin{equation}
    \frac{\partial f^\prime(\Gamma,t)}{\partial t} = -i\mcl{L}^\prime_{\mrm{c}}f^\prime(\Gamma,t) = -\left[ \frac{\partial}{\partial\Gamma}\cdot\dot{\Gamma}(t) + \dot{\Gamma}(t)\cdot\frac{\partial}{\partial\Gamma}\right]f^\prime(\Gamma,t),
    \label{eqn:Liouville_equation}
\end{equation}
where $\dot{\Gamma}$ are given by the equations of motion.
Since the equations of motion are derivable from a Hamiltonian, 
the condition of adiabatic incompressibility of phase space holds
so $\dot{\Gamma}\cdot\frac{\partial}{\partial\Gamma}=0$.
Separating the distribution function into  into its field-free 
and field-dependent parts $f^\prime(\Gamma,t)=f(\Gamma)+ \Delta f^\prime(\Gamma,t)$,
Eq.~\ref{eqn:Liouville_equation} becomes
\begin{equation}
     \frac{\partial }{\partial t}\left[f(\Gamma)+\Delta f^\prime(\Gamma, t)\right]= -\left[i\mcl{L}_{\mrm{c}} +i\Delta\mcl{L}^\prime_{\mrm{c}}(t) \right]\left[f(\Gamma)+ \Delta f^\prime(\Gamma,t)\right],
      \label{eqn:Liouville_equation2}
\end{equation}
The field-free part of the distribution function is 
the solution of the field-free Liouvillean $i\mcl{L}_{\mrm{c}}$
\begin{equation}
    \frac{\partial f(\Gamma)}{\partial t}  = -i\mcl{L}_{\mrm{c}}f(\Gamma) = 0.
    \label{eqn:field_independent}
\end{equation}
Upon substitution of Eq.~\ref{eqn:field_independent} in 
Eq.~\ref{eqn:Liouville_equation2} and keeping only first-order terms, the linearized Liouville equation follows 
\begin{equation}
        \frac{\partial  \Delta f^\prime(\Gamma, t)}{\partial t}  + i\mcl{L}_{\mrm{c}} \Delta f^\prime(\Gamma, t) = -i\Delta\mcl{L}^\prime_{\mrm{c}}(t)f(\Gamma).
        \label{eqn:Liouville_equation3}
\end{equation}
The solution of is Eq.~\ref{eqn:Liouville_equation3} is
\begin{equation}
     \Delta f^\prime(\Gamma, t) = -\int^t_0\!\mrm{d}t^\prime\, \mrm{e}^{i\mcl{L}_{\mrm{c}}(t-t^\prime)}\Delta i\mcl{L}^\prime_{\mrm{c}}(t^\prime) f (\Gamma).
     \label{eqn:solution}
\end{equation}
Using Eqs.~\ref{eqn:Liouville_equation}, \ref{eqn:Liouville_equation2} and \ref{eqn:field_independent}, the integrand can be written as
\begin{equation}
    \Delta i \mcl{L}^\prime_{\mrm{c}}(t)f(\Gamma) =i\mcl{L}^\prime_{\mrm{c}}(t)f_0(\Gamma)-i\mcl{L}_{\mrm{c}}(t)f(\Gamma) = \frac{\partial}{\partial\Gamma}\cdot\dot{\Gamma}(t)f(\Gamma)=\beta \dot{E}(t)f(\Gamma),
     \label{eqn:sub1}
\end{equation}
where the last equality relates  $\Delta i \mcl{L}^\prime_{\mrm{c}}$ to the adiabatic derivative of the internal energy $\dot{E}$.
For the perturbation considered, $\dot{E}$ is given as
\begin{equation}
    \dot{E} = - \overline{v}_x \sum^N_{i=1}\dot{p}_{ix}= - \overline{v}_x \sum^N_{i=1}f_{ix}= -\overline{v}_x F_x.
     \label{eqn:sub2}
\end{equation}
Substitution of  Eqs.~\ref{eqn:sub1} and \ref{eqn:sub2} 
in Eq.~\ref{eqn:solution} leads to
\begin{equation}
     \Delta f^\prime(\Gamma,t) = -\beta\int^t_0\!\mrm{d}t^\prime\, \mrm{e}^{i\mcl{L}_{\mrm{c}}(t-t^\prime)} F_x (\Gamma)\overline{v}_x f(\Gamma).
\end{equation}
With the change in distribution function under the perturbation obtained, we can write the linear response of the tangential force when the system has reached steady state as
\begin{equation}
   \langle F_x(t)\rangle^\prime_{\mrm{c}} - \langle F_x(0)\rangle_{\mrm{c}} =  \lim_{t\to\infty}\int\!\mrm{d}\Gamma\, F_x(\Gamma)\Delta f^\prime(\Gamma,t) = -\beta\int^\infty_0\!\mrm{d}t^\prime\int\!\mrm{d}\Gamma\,\mrm{e}^{i\mcl{L}_{\mrm{c}} t^\prime} F_x (\Gamma) F_x (\Gamma) \overline{v}_x f(\Gamma), 
\end{equation}
where $\langle\cdots\rangle_{\mrm{c}}^\prime$ denotes a canonical average in the 
perturbed constrained system. 
Using the fact that $F_x$ has zero mean at equilibrium, $\langle F_x(0)\rangle_{\mrm{c}}=0$, we arrive at
\begin{equation}
   \langle F_x(t)\rangle^\prime_{\mrm{c}} = -\beta \int^\infty_0\!\mrm{d}t\,\langle F_x(t) F_x(0) \rangle_{\mrm{c}} \overline{v}_x.
   \label{eq:linear_response}
\end{equation}
Combing Eqs.~\ref{eq:GK_friction} and  \ref{eq:linear_response} allows us to write the constitutive relation
\begin{equation}
    F_x = -\lambda_{\rm eff}\mcl{A}\overline{v}_x,
\end{equation}
which is valid in the linear response regime.


\subsection{Macroscopic hydrodynamics}

\begin{figure}[H]     \centering
  \includegraphics[width=0.95\linewidth]{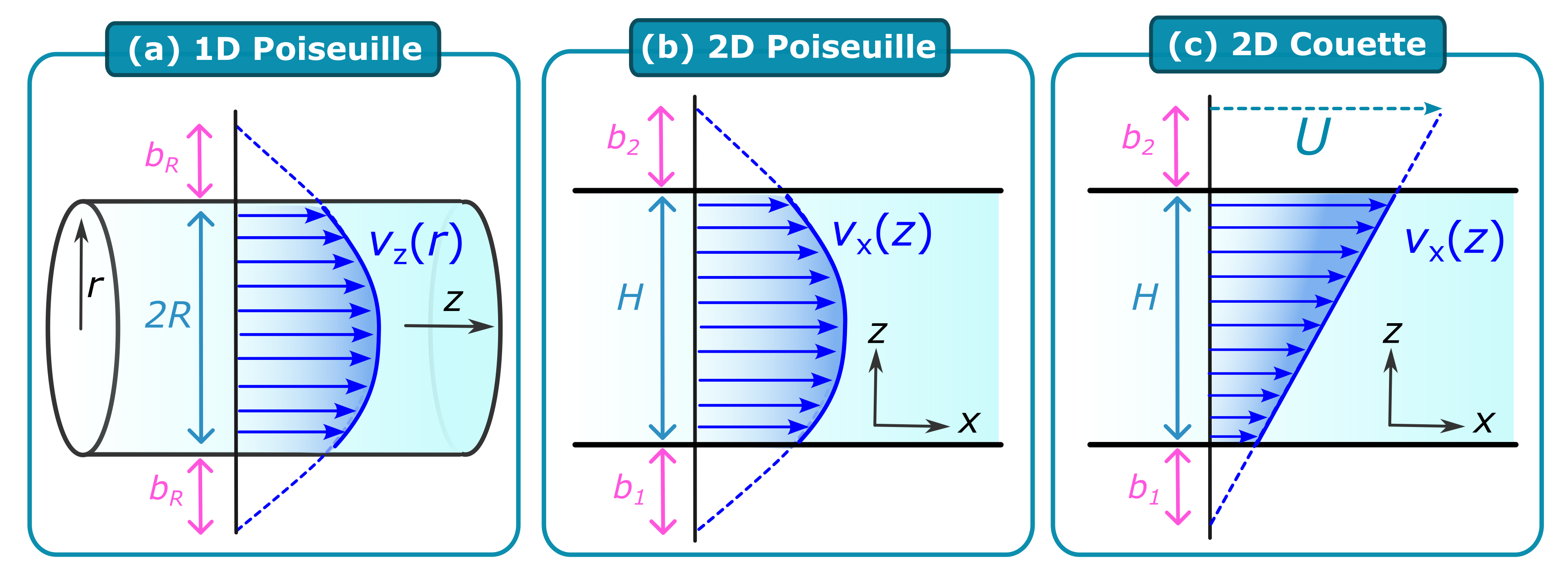}
  \caption{\textbf{Classical hydrodynamic problems considered.}
  The Navier--Stokes equation is solved for three cases including 
  (a) 1D Poiseuille flow of a pressure-driven fluid along a cylindrical tube, 
  (b) 2D Poiseuille flow of a pressure-driven fluid confined between two planes 
  and (c) 2D Couette flow of fluid confined between two planes sheared by the top plane moving at a constant velocity.
  The boundary conditions employed account for finite slippage
  at the liquid--solid interface.
  The slip length (pink) is defined as a distance from the boundary to where the linearly extrapolated fluid velocity profile (blue) vanishes. }
\label{fig:hydrodynamic_problem}
\end{figure}

In this section, the Navier--Stokes equation will be solved 
subject to the general slip boundary condition
for three cases: 1D Poiseuille flow, 2D Poiseuille flow
and 2D Couette flow, as shown schematically in Fig.~\ref{fig:hydrodynamic_problem}. 
These are all standard hydrodynamic problems found in textbooks
\cite{birdBook, LandauLifshitzBook} solved with the no-slip boundary
condition.

\subsubsection{1D Poiseuille flow} 

The system considered is a cylindrical tube
of radius $R$ and length $L_z$ with a curvature-dependent
slip length $b_R$ is shown schematically in 
Fig.~\ref{fig:hydrodynamic_problem}(a). The shear viscosity of 
the confined fluid is $\eta$ and the pressure difference between
the ends of the tube is $\Delta P$.
The Navier--Stokes equation for a pressure-driven steady
flow in the $z$ direction is given in  cylindrical
coordinates
\begin{equation}
   \eta  \frac{1}{r}\frac{\partial}{\partial r }\left(r \frac{\partial v_z(r)}{\partial r } \right) = -\frac{\Delta P}{L_z}.
\end{equation}
Through integration, we find the general solution
\begin{equation}
    v_z(r) = -\frac{\Delta P}{4 \eta L_z} r^2 + c_1 \ln r + c_2.
\end{equation}
The first constant of integration vanishes $c_1=0$ as the velocity must remain finite at the center of the pipe. 
The second constant of integration $c_2$ can be found by imposing the boundary condition at the wall of the tube
\begin{equation}
    \left.\frac{\partial v_z(r)}{\partial r}\right\rvert_{r=R}=-\frac{1}{b_R v_z(R)},
\end{equation}
where $R$ is radius of the tube and $b_R$ is the curvature-dependent slip length. The solution for the fluid velocity distribution is
\begin{equation}
    v_z(r) = \frac{\Delta P }{4 \eta L_z} (R^2 - r^2 + 2Rb_R).
\end{equation}
At the boundary, the slip velocity is given as
\begin{equation}
    v_z(R) =\frac{\Delta P }{2 \eta L_z}  R b_R.
    \label{eq:tube_slip_velocity}
\end{equation}
The mean fluid velocity average over the cross-section of the cylinder is given by
\begin{equation}
    \overline{v}_z = \frac{1}{\pi R^2}\int^R_0 \!\mrm{d}r\, 2\pi r v_z(r) = \frac{\Delta P}{8 \eta L_z}(R^2 + 4 R b_R).
    \label{eq:tube_velocity}
\end{equation}
By considering the shear stress of the fluid 
$\sigma_{r z}(r)$ at the wall $r=R$, the frictional force per area
is 
\begin{equation}
    \frac{F_z}{\mcl{A}} = \sigma_{r z}(R)= \eta\left.\frac{\partial v_z(r)}{\partial r}\right\rvert_{r=R} = -\frac{\Delta P }{2 L_z} R.
    \label{eq:tube_friction}
\end{equation}
Combining Eqs.~\ref{eq:tube_velocity} and \ref{eq:tube_friction}
with the constitutive relation
\begin{equation}
    F_z = -\lambda^{\mrm{1D}}_{\mrm{eff}}\mcl{A} \overline{v}_z,
\end{equation}
we arrive at an expression for the effective friction coefficient $\lambda^{\mrm{1D}}_{\mrm{eff}}$
\begin{equation}
    \lambda^{\mrm{1D}}_{\mrm{eff}}(R) = \frac{4\eta}{R + 4 b_R}.
    \label{eq:hydrodynamic_1D_eff}
\end{equation}
Upon rearrangement, the curvature-dependent slip length is given as
\begin{equation}
    b_R = \frac{\eta}{\lambda^{\mrm{1D}}_{\mrm{eff}}R)} - \frac{R}{4}. 
    \label{eqn:slip_length_tube}
\end{equation}
Combining Eqs.~\ref{eq:tube_slip_velocity} and \ref{eq:tube_friction}
with Navier's constitutive relation
\begin{equation}
    F_z = -\lambda^{\mrm{1D}}_{\mrm{intr}}\mcl{A} v_z(R),
\end{equation}
the intrinsic friction $\lambda^{\mrm{1D}}_{\mrm{intr}}$ is given as
\begin{equation}
    \lambda^{\mrm{1D}}_{\mrm{intr}} = \frac{\eta}{b_R} =\left(\frac{1}{\lambda^{\mrm{1D}}_{\mrm{eff}}(R)}-\frac{R}{4\eta}\right)^{-1}.
\end{equation}
Note that the intrinsic and effective friction coefficients are only the same
$\lambda^{\mrm{1D}}_{\mrm{eff}} =\lambda^{\mrm{1D}}_{\mrm{intr}}$
in the limit of small tube and large slippage, $R \lambda^{\mrm{1D}}_{\mrm{eff}}(R) \ll 4\eta$.

\subsubsection{2D Poiseuille flow}

The system considered is a two-dimensional channel
made up of two parallel plates lying in the $xy$-plane,
as shown schematically in Fig.~\ref{fig:hydrodynamic_problem}(b).
This distance between the planes is given by $H$ while 
$b_1$ and $b_2$ are the slip lengths of the lower plane at $z=0$ 
and the upper plane at $z=H$ respectively.
The confined fluid has a shear viscosity as $\eta$ 
and  $\Delta P$ is now the pressure difference between two
ends of the channel across a width of $L_x$.
The Navier--Stokes equation for a pressure-driven steady 
flow of fluid along $x$ is given in Cartesian coordinates
\begin{equation}
        \eta \frac{\partial^2 v_x(z)}{\partial z^2} =  -\frac{\Delta P}{L_x}.
\end{equation}
Through integration, the general solution obtained is
\begin{equation}
    v_x(z) = -\frac{1}{2\eta}\frac{\Delta P}{L_x}z^2 + c_1 z + c_2. 
\end{equation}
To determine the constants of integration $c_1$ and $c_2$, we
employ the boundary conditions
\begin{equation}
    \left.\frac{\partial v_x(z)}{\partial z}\right\rvert_{z=0}=\frac{1}{b_1}v_x(0), \quad\quad \left.\frac{\partial v_x(z)}{\partial z}\right\rvert_{z=H}=-\frac{1}{b_2}v_x(H).
\end{equation}
The solution for the flow profile is
\begin{equation}
    v_{x}(z)=\left[-z^2 +\frac{(z + b_1)(H+2b_2)H}{H+b_1 + b_2}\right]\frac{1}{2\eta}\frac{\Delta P}{L_x}.
    \label{eq:full_solution}
\end{equation}
At the boundaries, we have the slip velocities as 
\begin{equation}
\begin{split}
    v_{x}(0) &= \frac{(H+2b_2)H }{H+b_1 + b_2}\frac{b_1}{2\eta}\frac{\Delta P}{L_x}, \\
    v_{x}(H) &=  \frac{(H+2b_1)H }{H+b_1 + b_2}\frac{b_2}{2\eta}\frac{\Delta P}{L_x}.
\end{split}
\end{equation}
The mean fluid velocity average over the height of the channel is given by
\begin{equation}
    \overline{v}_x=\frac{1}{H}\int^H_0 \!\mrm{d}z\,v_x(z)=\frac{[H^2 + 4 H(b_1 + b_2)+12b_1 b_2]H}{H + b_1 + b_2}\frac{1}{12\eta}\frac{\Delta P}{L_x}.
    \label{eq:mean_velocity_channel}
\end{equation}
The frictional force per area from each wall on the fluid is given as
\begin{equation}
\begin{split}
      \frac{F_{x,1}}{\mcl{A}} &=  \sigma_{xy}(0)=-\eta \left.\frac{\partial v_x(z)}{\partial z}\right\rvert_{z=0} = -\frac{\eta}{b_1} v_x(0)=-\frac{(H+2b_2)H}{H + b_1 + b_2}\frac{1}{2}\frac{\Delta P}{L_x},\\
      \frac{F_{x,2}}{\mcl{A}} &=     \sigma_{xy}(H)=\eta \left.\frac{\partial v_x(z)}{\partial z}\right\rvert_{z=H}= -\frac{\eta}{b_2} v_x(H)=-\frac{(H+2b_1)H}{H + b_1 + b_2}\frac{1}{2}\frac{\Delta P}{L_x}.
\end{split}
\end{equation}
The total tangential frictional force per surface area on the fluid from the wall is therefore
\begin{equation}
    \frac{F_x}{\mcl{A}} = \frac{F_{x,1}+F_{x,2}}{\mcl{A}}   =-\frac{ H \Delta P}{L_x}.
    \label{eq:channel_friction}
\end{equation}
Combining Eqs.~\ref{eq:mean_velocity_channel} and \ref{eq:channel_friction}
with the constitutive relation
\begin{equation}
    F_x = -\lambda^{\mrm{2D}}_{\mrm{eff}}\mcl{A} \overline{v}_x,
    \label{eq:constitutive_relation}
\end{equation}
we arrive at an expression for the effective friction coefficient $\lambda^{\mrm{2D}}_{\mrm{eff}}$
\begin{equation}
    \lambda^{\mrm{2D}}_{\mrm{eff}}(H) = \frac{12(H + b_1 + b_2)\eta}{H^2 + 4 H(b_1 + b_2)+12b_1 b_2}.
    \label{eq:friction_eff_2D}
\end{equation}
In comparison, by considering Navier's constitutive relations
\begin{equation}
    F_{x,1} = - \lambda^{\mrm{2D}}_{\mrm{intr},1}\mcl{A}v_x(0), \quad\quad F_{x,2} = - \lambda^{\mrm{2D}}_{\mrm{intr},2}\mcl{A}v_x(H),
\end{equation}
the intrinsic friction coefficients are
\begin{equation}
    \lambda^{\mrm{2D}}_{\mrm{intr},1}=\frac{\eta}{b_1}, \quad\quad \lambda^{\mrm{2D}}_{\mrm{intr},2}=\frac{\eta}{b_2}.
\end{equation}
Before going further, it is helpful to look at some limits of Eq.~\ref{eq:friction_eff_2D}. In the limit that the channel has a very large height or no slippage at both walls
\begin{equation}
  \lim_{H\to\infty}\lambda^{\mrm{2D}}_{\mrm{eff}}(H)  = \lim_{b_1, b_2 \to 0}\lambda^{\mrm{2D}}_{\mrm{eff}}(H)  = \frac{12\eta}{H}. 
   \label{eqn:limit_only_viscosity}
\end{equation}
In the limit of small channel
\begin{equation}
   \lim_{H=0}\lambda^{\mrm{2D}}_{\mrm{eff}}(H) = \frac{\eta}{b_1}+\frac{\eta}{b_2}  = \lambda^{\mrm{2D}}_{\mrm{intr},1}+\lambda^{\mrm{2D}}_{\mrm{intr},2}.
\end{equation}
In the limit where friction is dominated by one side of the channel such that $b_2\gg b_1$ and $b_2\gg H$ 
\begin{equation}
   \lim_{b_2\rightarrow \infty}\lambda^{\mrm{2D}}_{\mrm{eff}}(H)   = \frac{3 \eta}{H + 3 b_1}=\left(\frac{H}{3\eta}+\frac{1}{\lambda^{\mrm{2D}}_{\mrm{intr},1}}\right)^{-1}.
\end{equation}
We expect that such a relation can describe friction at an open
water--air interface.

To obtain a closed expression relating the slip length and the effective friction coefficient, we can consider the case of a 
symmetric channel $b_1=b_2=b$ such that
\begin{equation}
\lambda^{\mrm{2D}}_{\mrm{eff}}(H) = \frac{12(H + 2 b)\eta}{H^2 + 8 H b+12 b^2 }. 
\end{equation}
By rearranging  $b$ and taking the positive solution, we arrive at
\begin{equation}  
 b= \left(\frac{\eta}{\lambda^{\mrm{2D}}_{\mrm{eff}}(H)} - \frac{H}{3}\right) + \left[\left(\frac{\eta}{\lambda^{\mrm{2D}}_{\mrm{eff}}(H)} - \frac{H}{3}\right)^2 + H \left(\frac{\eta}{\lambda^{\mrm{2D}}_{\mrm{eff}}(H)} - \frac{H}{12}\right)\right]^{1/2}.
 \label{eqn:slip_length}
\end{equation}

\subsubsection{2D Couette flow}

The system considered is a two-dimensional channel
made up of two parallel plates lying in the $xy$-plane,
as shown schematically in Fig.~\ref{fig:hydrodynamic_problem}(c).
This distance between the planes is given by $H$ while 
$b_1$ and $b_2$ are the slip lengths of the lower plane at $z=0$ 
and the upper plane at $z=H$ respectively.
The confined fluid has a shear viscosity as $\eta$. We suppose that the upper plate is moving with a constant
velocity $U$ while the bottom plane is stationary in the laboratory
frame of reference.
The Navier--Stokes equation for a shear-driven steady 
flow of fluid along $x$ is given in Cartesian coordinates
\begin{equation}
    \frac{\partial^2 v_x(z)}{\partial z^2} = 0.
\end{equation}
Upon integration, we find the general solution
\begin{equation}
    v_x(z) = c_1 z + c_2.
\end{equation}
To determine the constants of integration $c_1$ and $c_2$,
we use the following 
boundary conditions
\begin{equation}
    \frac{\partial v_x(z)}{\partial z}=\frac{1}{H+ b_1 + b_2}U=\frac{1}{b_1}v_x(0),
\end{equation}
specifying the gradient of the velocity profile and 
the slip velocity at the upper wall.
The velocity profile of the fluid in the laboratory 
frame of reference is therefore
\begin{equation}
    v_{x}(z)=\left(z+b_1\right) \frac{U}{H+b_1+b_2}.
    \label{eq:solution_Couette}
\end{equation}
The velocities at the boundaries are
\begin{equation}
\begin{split}
    v_{x}(0) &= \frac{b_1}{H+b_1+b_2} U, \\
    v_{x}(H) &= \frac{H+b_1}{H+b_1+b_2}U.
\end{split}
\end{equation}
The mean fluid velocity average over the height of the channel is given by
\begin{equation}
    \overline{v}_x=\frac{H+2b_1}{H+b_1+b_2}\frac{U}{2}.
\end{equation}

The frictional force per area from each wall on the fluid is given as
\begin{equation}
\begin{split}
      \frac{F_{x,1}}{\mcl{A}} &=  \sigma_{xy}(0)=-\eta \left.\frac{\partial v_x(z)}{\partial z}\right\rvert_{z=0} = -\eta\frac{U}{H+b_1+b_2},\\
      \frac{F_{x,2}}{\mcl{A}} &=     \sigma_{xy}(H)=\eta \left.\frac{\partial v_x(z)}{\partial z}\right\rvert_{z=H}=\eta\frac{U}{H+b_1+b_2}.
\end{split}
\end{equation}
Here, we can see clearly that the fluid will experience equal and opposite forces  from each wall, so the total frictional force vanishes
\begin{equation}
    F_x=F_{x,1}+F_{x,2}=0.
\end{equation}
Moreover, since both plates are moving at different velocities, it is not possible to talk about the mean velocity of the fluid in
a single frame of reference of the solid and therefore an effective friction coefficient. The constitutive relation in Eq.~\ref{eq:constitutive_relation} does not apply in the case of Couette flow.  
It is possible to use Navier's constitutive relation using the slip velocity defined relative to the adjacent wall
\begin{equation}
\begin{split}
    v_{x,1}(0) &= v_x(0) - 0 =  \frac{b_1}{H+b_1+b_2} U, \\
    v_{x,2}(H) &= v_x(H) - U = -\frac{b_2}{H+b_1+b_2}U.
\end{split}
\end{equation}
By writing 
\begin{equation}
    F_{x,1} = - \lambda^{\mrm{2D}}_{\mrm{intr},1}\mcl{A}v_{x,1}(0), \quad\quad F_{x,2} = - \lambda^{\mrm{2D}}_{\mrm{intr},2}\mcl{A}v_{x,2}(H),
\end{equation}
we can check for consistency that the same intrinsic coefficients as in the case of 2D Poiseuille flow 
\begin{equation}
    \lambda^{\mrm{2D}}_{\mrm{intr},1}=\frac{\eta}{b_1}, \quad\quad \lambda^{\mrm{2D}}_{\mrm{intr},2}=\frac{\eta}{b_2},
\end{equation}
are obtained.

\newpage

\section{Verification with molecular simulations}
\label{sec:verification}

\subsection{System set-up}

The systems considered have water confined between
two solid substrates of different attractive strengths.
Four symmetric channels are made up of two substrates 
with the same wall--fluid attraction $\varepsilon_{\rm wf}=\alpha \varepsilon_0$,
where $\varepsilon_0=1.57\,\mrm{kJ\,mol^{-1}}$ and the ``wetting coefficients" are $\alpha=1,2,3,4$. 
One asymmetric channel with 
$\alpha=1$ for the top wall and
$\alpha=4$ for the bottom wall was also considered.
Various channel heights are considered
including $H/ \mrm{nm}\approx 1.4, 2.7, 5.2$
between the first atomic planes of the solids for
$N=1080, 2160, 3240$ water molecules, respectively.
In addition, we also considered some systems with fewer number of water molecules $N=233,333,498$ for the sub-nanometric confinement regime.
Each solid substrate consists of 1620 atoms fixed on an face-centered-cubic lattice ($9\times9\times5$ unit cells with a lattice
parameter of $0.407\,\mrm{nm}$ corresponding to the unit cell of gold), resulting in ten atomic
planes perpendicular to the $z$ direction and facing the inner part of the system with a (100) plane. 
For all the systems, the orthogonal simulation box has
lateral dimensions $\ell_x = \ell_y = 3.663\,\mrm{nm}$ and $\ell_z$ is chosen to be large enough to minimise the slab interaction with its periodic images.
The initial configuration for the smallest system was
obtained from Ref.~\onlinecite{Pireddu2023}.
We summarise the systems considered in Table.~\ref{tab:2d_LJ_channels}. 

\begin{table}[H]
    \centering
    \begin{tabular}{c  c c  c}
    \hline
    \hline
    $\alpha$ &  $H\,[\mrm{nm}]$  & $N$   \\
    \hline
    1 & 7.825 &  3240 \\
    1 & 5.435 &  2160 \\
    1 & 2.956 &  1080 \\
    1 & 1.641 & 498 \\
    1 & 1.269 & 332 \\
    \hline
    2 & 7.694 &  3240 \\
    2 & 5.255 &  2160 \\
    2 & 2.825 &  1080 \\
    2 & 1.513 &  498 \\
    2 & 1.139 &  332 \\
     \hline
  3 &   7.621 & 3240 \\
  3 &   5.181 & 2160 \\
  3 &   2.750 & 1080 \\
  3 &   1.440 & 498 \\
  3 &   1.064 & 332 \\
  3 &   0.859 & 233 \\
   \hline
    4  & 7.567 &  3240 \\
    4  & 5.131 & 2160 \\
    4  & 2.695 &  1080 \\
    4  & 1.385 &  498 \\
    4  & 0.993 &  332 \\
    4  & 0.842 &  233 \\
     \hline
    1, 4  & 2.956 &  1080 \\
    1, 4  & 1.526 &  498 \\
    \hline
    \hline
    \end{tabular}
\caption
{
\textbf{Systems considered in molecular simulations of 2D flow.} 
Water is confined between two Lennard--Jones solid substrates. 
For each system, we report the wetting coefficient $\alpha$, 
the equilibrium height $H$
and the number of confined water molecules $N$.
}
\label{tab:2d_LJ_channels}
\end{table}

\subsection{Equilibrium molecular dynamics  details}

All simulations were carried out with the \texttt{LAMMPS}
simulations package \cite{Thompson2022}. 
Water--water interactions were described with the SPC/E 
water model \cite{Berendsen1987}. 
The geometry of water molecules was constrained using 
the \texttt{RATTLE} algorithm \cite{Andersen1983}.
The atoms on the solid substrates were held rigid with respect to its lattice. Each oxygen atom on a water molecule interacts with
a substrate atom through a 12-6 Lennard-Jones potential 
with a well-depth of $\epsilon_{\mrm{wf}}$ as quoted
in the previous section and $\sigma_{\mrm{wf}}=0.3059\,\mrm{nm}$.
All Lennard-Jones interactions were truncated and shifted at 
$1\,\mrm{nm}$. 
Electrostatic interactions were evaluated in real space up to
$1\,\mrm{nm}$ and long-ranged interactions were evaluated
using particle--particle particle--mesh Ewald summation
\cite{hockney1988} such that
the RMS error in the forces was a factor of $10^5$ smaller 
than the force between two unit charges separated by a 
distance of $0.1\,\mrm{nm}$ \cite{Kolafa1992}.

The simulations were carried out in the canonical 
(NVT) ensemble. The temperature of liquid water
is maintained at $T=298\,\mrm{K}$ using a
Nos\'{e}--Hoover chain with 5 thermostats and a damping constant of $\SI{0.1}{\pico\second}$.
Dynamics were propagated using the velocity Verlet 
algorithm with 
time-step of $\SI{1}{\femto\second}$.
Each simulation box is prepared with a preliminary
equilibration of $1\,\mrm{ns}$ where an additional force in the
$z$-direction corresponding to a pressure of $1\,\mrm{atm}$ was applied to the solid substrate at the top while the solid substrate
at the bottom was fixed. At this stage, the substrates can move
vertically and act as pistons and the equilibrium channel height $H$
is determined as an average. 
Then, the positions of the solid substrate were held fixed and an
equilibration run of $1\,\mrm{ns}$ was performed in the conditions used for production. The production runs for each system 
are $10\,\mrm{ns}$ long.
During the production, the total summed force acting on all
water molecules in both $x$ and $y$ directions were sampled 
at each time-step.
For the constrained EMD simulations, the linear momentum is zeroed
in the $x$ and $y$ directions by substracting the center-of-mass
velocity of the water at every time-step using the \texttt{fix momentum} command in \texttt{LAMMPS}.
Simulation input scripts are made available at Ref.~\onlinecite{Bui2024}.

\subsection{Non-equilibrium molecular dynamics  details} 

For the same systems considered above, NEMD simulations 
for a pressure-driven flow were carried out as reference.
To mimic the effect of a pressure gradient, an individual 
external force is applied on the oxygen
atom of each water molecules in the $x$ direction
while keeping the walls immobile, generating a 2D
Poiseuille flow.
The thermostat was applied to the water
only after excluding the center-of-mass contribution.
For each system considered, 
at least seven different simulations were run with different
magnitudes of the total external force $F^{\mrm{ext}}_x$. 
The external force magnitudes were chosen such that the resulting velocities
of the fluid do not exceed $50\,\mrm{m s}^{-1}$ to maintain
the linear response between the frictional force and the velocity.
Using the configuration from EMD simulations, an equilibration NEMD run of $200\,\mrm{ps}$ was performed, followed by $10\,\mrm{ns}$
of production run.
During the production, the total force from the solid on the water $F_x$, the total fluid velocity  $\overline{v}_x$ and the velocity profile   $\overline{v}_x(z)$ were sampled.

In addition to Poiseuille flow, NEMD simulations for a shear-driven flow were also carried out. Here, 
the top substrate was moved at a constant velocity $U$ in the $x$ direction while the bottom substrate was held fixed, generating a 2D Couette flow.
Similarly, the thermostat was applied to the water
only after excluding the center-of-mass contribution.
For each system, the magnitudes of the top wall
velocity $U$ were chosen such that the resulting velocities
of the fluid do not exceed $50\,\mrm{m s}^{-1}$.
Using the configuration from EMD simulations, an equilibration NEMD run of $200\,\mrm{ps}$ was performed, followed by $10\,\mrm{ns}$
of production run.
During the production, 
the velocity profile of the fluid $\overline{v}_x(z)$ and 
the force from each wall on the fluid $F_{x,1}$ and $F_{x,2}$  were sampled.

\subsection{Green--Kubo relation for the effective friction} 

\begin{figure}[H]     \centering
  \includegraphics[width=0.95\linewidth]{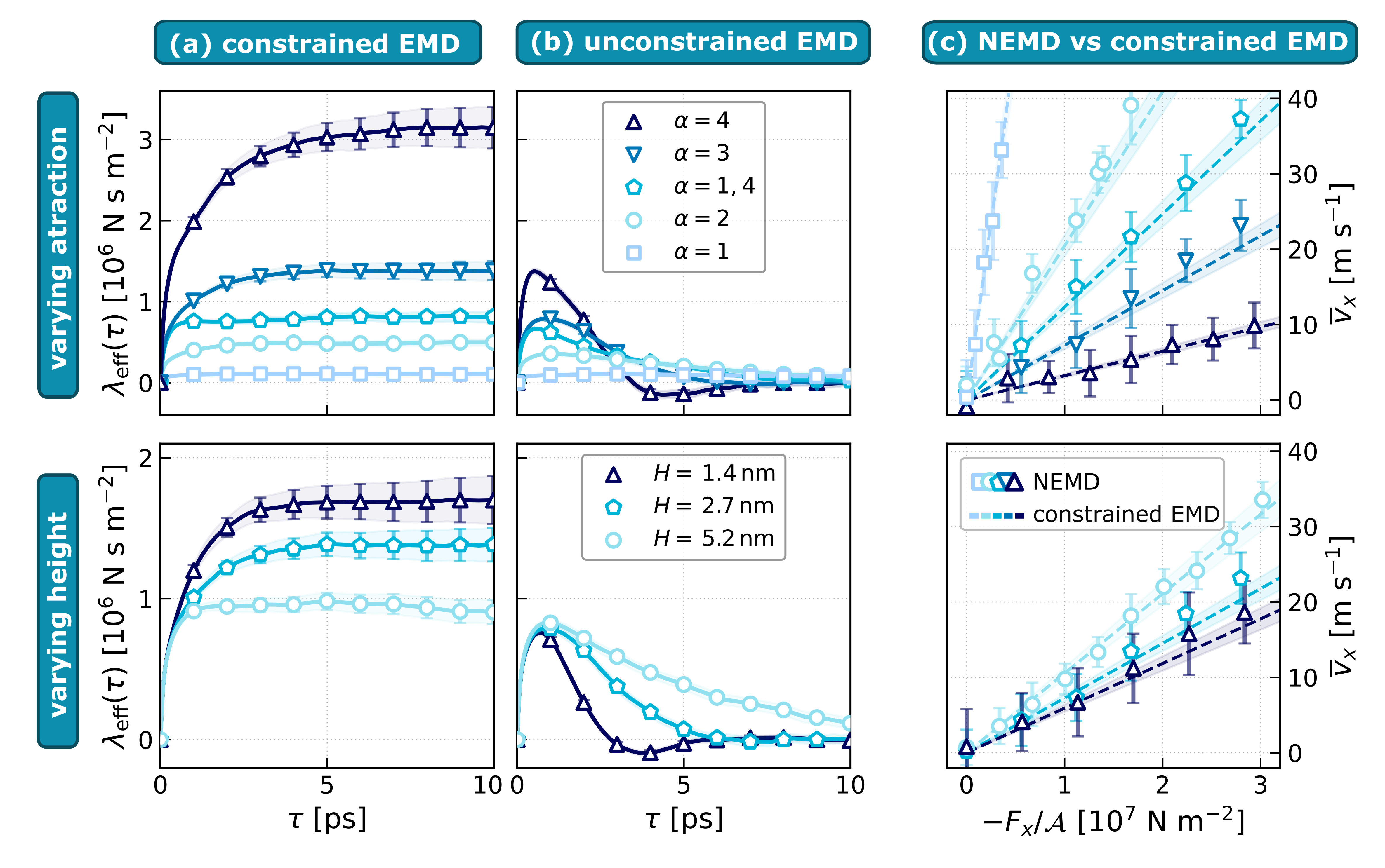}
  \caption{\textbf{Effective friction from EMD and NEMD simulations.} (a) In EMD simulations with the constraint zeroing the liquid momentum, the Green--Kubo integral in Eq.~\ref{eq:GK_friction} gives a well-defined plateau as
  a measurement of the effective friction $\lambda_{\mrm{eff}}$. (b) When no constraint is applied, the integral decays to zero at long time. Such a plateau problem is more severe for channels with
  larger attraction and of smaller height. (c) In the NEMD simulations, the average fluid velocity $\overline{v}_x$ of  Poiseuille flow is shown as a function of the frictional force per unit area as data points. The dashed lines are corresponding predictions of linear response from the constitutive relation in Eq.~\ref{eq:constitutive_relation} using $\lambda_{\mrm{eff}}$ determined from the constrained EMD simulation.
  The top panel shows the results for channels with different wall attractive strengths with $H\approx2.7\,\mrm{nm}$. 
  The bottom panel shows the results for channels of different heights for the  system where both walls have $\alpha = 3$.}
\label{fig:green_kubo_all_channels} 
\end{figure}

The Green--Kubo friction computed from EMD simulations
as the integral of the force autocorrelation function
with and without the zero momentum constraint for various systems 
considered with different attractive strengths and 
channel heights are shown in 
Figs.~\ref{fig:green_kubo_all_channels}(a) and (b). 
The error bars correspond to the
standard errors with a 95\% confidence interval
obtained from splitting the  entire trajectory 
into 100 blocks such that each block 
is $100\,\mrm{ps}$ long. While the Green--Kubo integral 
from the unconstrained EMD simulation decays to zero
at longer time, it reaches a well-defined plateau in the
constrained EMD simulations, which is the effective
friction $\lambda_{\mrm{eff}}$ of fluid flow in the channel.
In Fig.~\ref{fig:green_kubo_all_channels}(c), the constitutive relation between the frictional force and mean fluid velocity in Eq.~\ref{eq:constitutive_relation} is verified for channels of different wall attractions and heights. The NEMD results are from 2D Poiseuille flow simulations. We verified that the total friction force on the fluid on average is equal and opposite to the external force applied
\begin{equation}
    \langle F_x \rangle = -F_x^{\mrm{ext}},
\end{equation}
justifying using the external force when plotting the force-flux relation. 
From the constrained EMD simulations, and $\overline{v}_x$ is plotted as a function of $-F_x/\mcl{A}$ as expected from using $\lambda_{\mrm{eff}}$ in the constitutive relation
in Eq.~\ref{eq:constitutive_relation}.

\subsection{Fluctuation--dissipation theorem}

\begin{figure}[H]     \centering
  \includegraphics[width=0.95\linewidth]{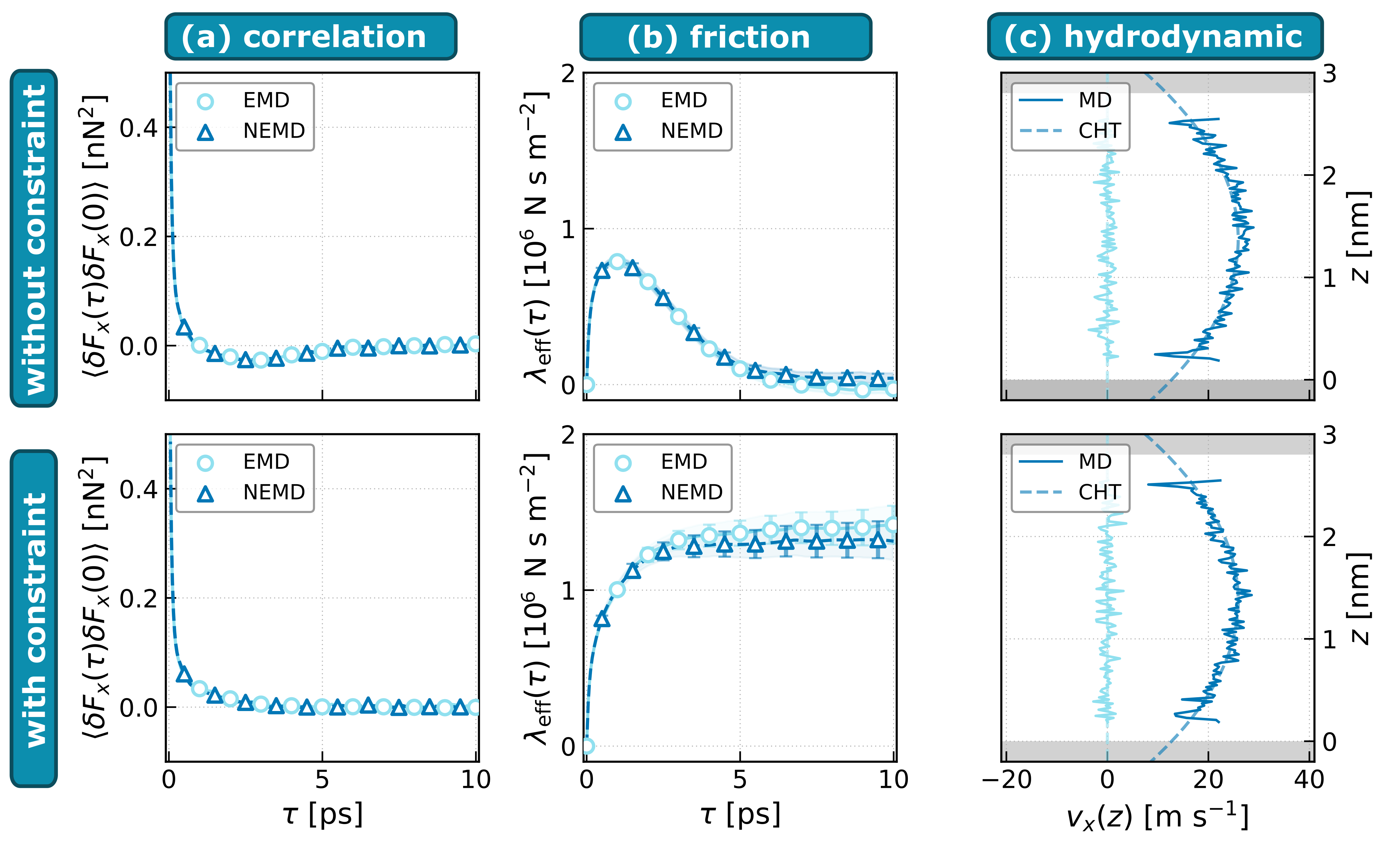}
  \caption{\textbf{The frictional response and  hydrodynamic boundaries in constrained and unconstrained simulations.} 
  We show results of unconstrained dynamics in the top panel and of dynamics under zero liquid momentum constraint in the bottom panel. In the NEMD simulation, a 2D Poiseuille flow is simulated with an average fluid velocity of $22\,\mrm{m\,s^{-1}}$, when the frictional response of the system remains linear. Both in and out-of-equilibrium,  the (a) force
  correlation and therefore (b) its Green--Kubo friction integral
  remains the same, verifying that the dissipation--fluctuation theorem holds. While the constraint placed on the momentum of the liquid helps evaluating the effective friction coefficient without the plateau problem, it does not change the 
  overall hydrodynamic boundary of the system. (c) With or 
  without such a constraint, the velocity profiles
  from MD simulations both agree well with the CHT prediction from  Eq.~\ref{eq:full_solution}.
  }
\label{fig:FDT_check}
\end{figure}

When the system responses linearly, we expect the correlation
in the frictional force of the system is the same in equilibrium
and out-of-equilibrium. Specifically, the effective friction
remains unchanged when the system is close enough to equilibrium such that it can be computed in NEMD simulations using
\begin{equation}
    \lambda^\prime_{\rm eff} = \frac{\beta}{\mcl{A}}\int^\infty_0\!\mrm{d}\tau\,\langle  \delta F_x(\tau) \delta F_x(0)  \rangle^\prime_{\mrm{c}},
\end{equation}
where the  Green--Kubo integral is now defined with the correlation of the deviation of the tangential force away from the average 
\begin{equation}
    \delta F_x(t) = F_x(t) - \langle F_x \rangle^\prime_{\mrm{c}}.
\end{equation}
To verify that the fluctuation--dissipation theorem holds within the linear response regime, we compare the 
friction response of the system 
in and out of equilibrium for a representative case in Figs.~\ref{fig:FDT_check}(a) and (b).
The force autocorrelation and friction integral agree between unconstrained EMD and NEMD Poiseuille flow simulations, both suffering from the plateau problem.
When a constraint is placed on the momentum of the liquid, i.e
\begin{equation}
    P_x = 0\quad\quad \mrm{[EMD]}, \quad\quad\quad
    P_x =M\overline{v}_x \quad\quad \mrm{[NEMD]},
\end{equation}
the force autocorrelation and friction integral agree between constrained EMD and NEMD simulations.
In practice, we made use of change of frame of reference to perform the NEMD constrained simulation. First the average fluid velocity $\overline{v}_x$ is determined from an unconstrained NEMD.
Then in the constrained NEMD simulation with the same external force applied to the fluid, both solid substrates are moved at a constant velocity of $\overline{v}_x$ and the momentum of the liquid is zeroed at every time-step. In this way, the  mean fluid velocity 
is zero in the simulation frame of reference but is constrained at $\overline{v}_x$ in the solid frame of reference.

\subsection{Hydrodynamic boundary unaffected by constraints}

In Fig.~\ref{fig:FDT_check}(c), we compare the Poiseuille flow velocity profiles from 
an unconstrained and constrained NEMD simulations. We note here that the results are shown in the solid frame of reference. The agreement of the velocity profile between the  simulations with and without constraint confirms that the
hydrodynamic boundary is unaffected by the momentum conservation constraint.
For reference, we also show the velocity profiles expected from CHT in Eq.~\ref{eq:full_solution}
using the slip length determined in the next section.

\subsection{Slip length and intrinsic friction prediction} \label{sec:sliplength}

In Fig.~2(b) of the main article, we compare the intrinsic 
frictional properties for the liquid--solid interface formed 
between water and the solid substrates with $\alpha=1,2,3,4$
computed using four different methods. For all methods,
we have shown results using simulations of each symmetric channel with $N=1080$.

From the reference NEMD simulations of Poiseuille flow in each channel,  we obtained the average fluid velocity $\langle \overline{v}_x \rangle$ as a function of the external
driving force $F_{\mrm{ext}}$. 
Then, the effective friction $\lambda^{\mrm{2D}}_{\mrm{eff}}$ is computed as the following ratio $F_{\mrm{ext}}/(\mcl{A}\langle \overline{v}_x \rangle)$ based on the constitutive relation in Eq.~\ref{eq:constitutive_relation}. 
Then the slip length $b$ from NEMD is computed from Eq.~\ref{eqn:slip_length_sym}, with
the error estimated by propagation of statistical error on $\langle \overline{v}_x \rangle$.
We note that while in principle, 
the slip length can be obtained by direct extrapolation of the velocity profile from from NEMD simulations, this approach is subject to 
a high degree of uncertainty \cite{Kannam2012}.

From the EMD simulations with the zero momentum constraint on the liquid, $\lambda^{\mrm{2D}}_{\mrm{eff}}$ can be obtained as the
well-defined plateau of the Green--Kubo integral of the force autocorrelation function following Eq.~\ref{eq:GK_friction}. In practice, the plateau value is taken at  the long time limit ($t\approx 10\,\mrm{ps}$) of the Green--Kubo integral.
Again,  $b$ is then  computed based on Eq.~\ref{eq:constitutive_relation}, with
the error estimated by propagation of statistical error on $\lambda^{\mrm{2D}}_{\mrm{eff}}$. We refer to this approach as 'GK+CHT'.

For comparison, we also employed Bocquet and Barrat's formula \cite{Bocquet1994, Bocquet2013} (BB) to compute the intrinsic friction $\lambda_{\mrm{intr}}$
from unconstrained EMD simulations
\begin{equation}
   \lambda_{\mrm{BB}}(\tau) = \frac{\beta}{\mcl{A}} \int^{\tau}_0\!\mrm{d}t\,\langle F_x(t) F_x(0) \rangle.
    \label{eqn:BB}
\end{equation}
According to Refs.~\onlinecite{Bocquet1994, Bocquet2013}, the intrinsic 
friction would be given as the long time limit $\lambda_{\mrm{BB}}(\tau\to\infty)$,
which in practice we use $\tau\approx 10\,\mrm{ps}$.
Since, in general, there is no well-defined plateau value for the BB integral as it decays to zero at long time, a practical approach  often used in the literature is to take the maximum value of the integral before it starts to decay $\mrm{max}[\lambda_{\mrm{BB}}]$.

In each case, the slip length can be converted to the intrinsic friction coefficient via
\begin{equation}
    b = \frac{\eta}{\lambda_{\mrm{intr}}}.
\end{equation}
Here, we use a literature\cite{Gonzalez2010} value for the shear viscosity of bulk SPC/E water at $298\,$K as $\eta=0.729\,\mrm{mPa\, s}$. 


\subsection{Verifying velocity profile prediction}

\begin{figure}[H]     \centering
  \includegraphics[width=0.85\linewidth]{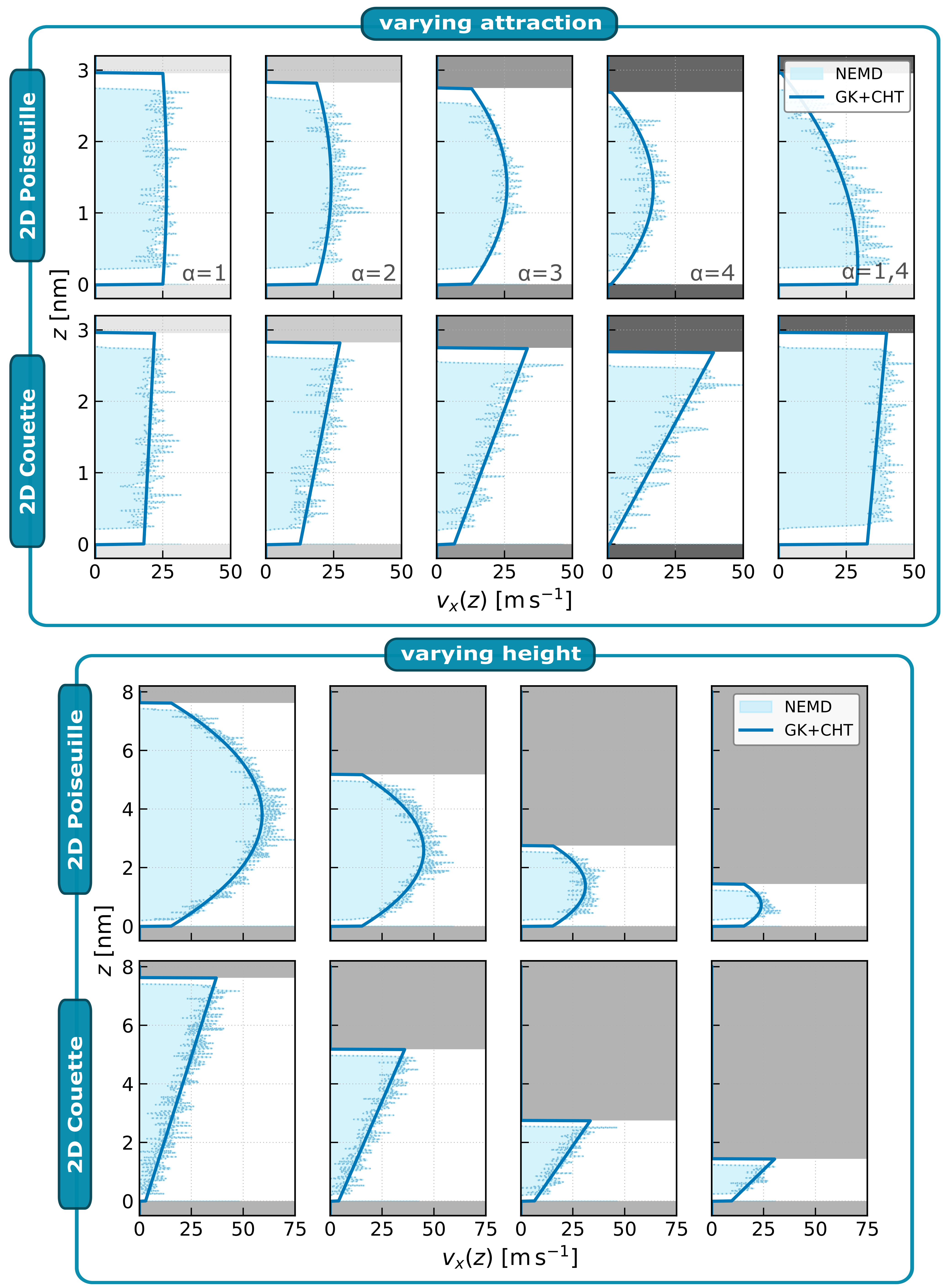}
  \caption{\textbf{ Prediction of
 velocity profiles of fluid flow.} 
    Comparison between the predicted velocity profiles from
  classical hydrodynamic theory (CHT) with boundary determined with
  the Green--Kubo (GK) relation and direct NEMD data are shown for 2D Poiseuille and 2D Couette flow.
  In the top panel, results for the symmetric channels with wetting coefficients
  $\alpha=1,2,3,4$ and asymmetric channel $\alpha=1,4$ with similar heights $H\approx2.7\,\mrm{nm}$ are shown. The solid walls are shaded darker for higher wall--liquid attractive strength. In the bottom panel,
  results are shown for various channel heights $H/\mrm{nm}=7.6,5.2,2.7,1.4$ are shown for the case of the symmetric $\alpha=3$ channel. In all cases, the velocity profiles predicted are in excellent agreement with NEMD data.
  }
\label{fig:EMD_vs_NEMD_attraction}
\end{figure}

With the slip length of the individual surfaces obtained from the constrained EMD simulations, 
we can also compare directly the prediction of the velocity profile using the solution to the Navier--Stokes equation
and the reference NEMD simulations. In Fig.~\ref{fig:EMD_vs_NEMD_attraction}, we show these
results for both 2D Poiseuille flow and 2D Couette flow using Eq.~\ref{eq:full_solution} and Eq.~\ref{eq:solution_Couette} respectively.
In the case of Poiseuille flow, the pressure gradient in the simulation is $-\Delta P/L_x=-F_\mrm{ext}/(H L_x L_y )$. 
Representative different channels are considered with varying wall--fluid attraction and varying height.

\newpage

\section{Sensitivity to hydrodynamic boundary position} \label{sec:sensitivity_theory}

\subsection{Theory}

\begin{figure}[H]     \centering
  \includegraphics[width=0.82\linewidth]{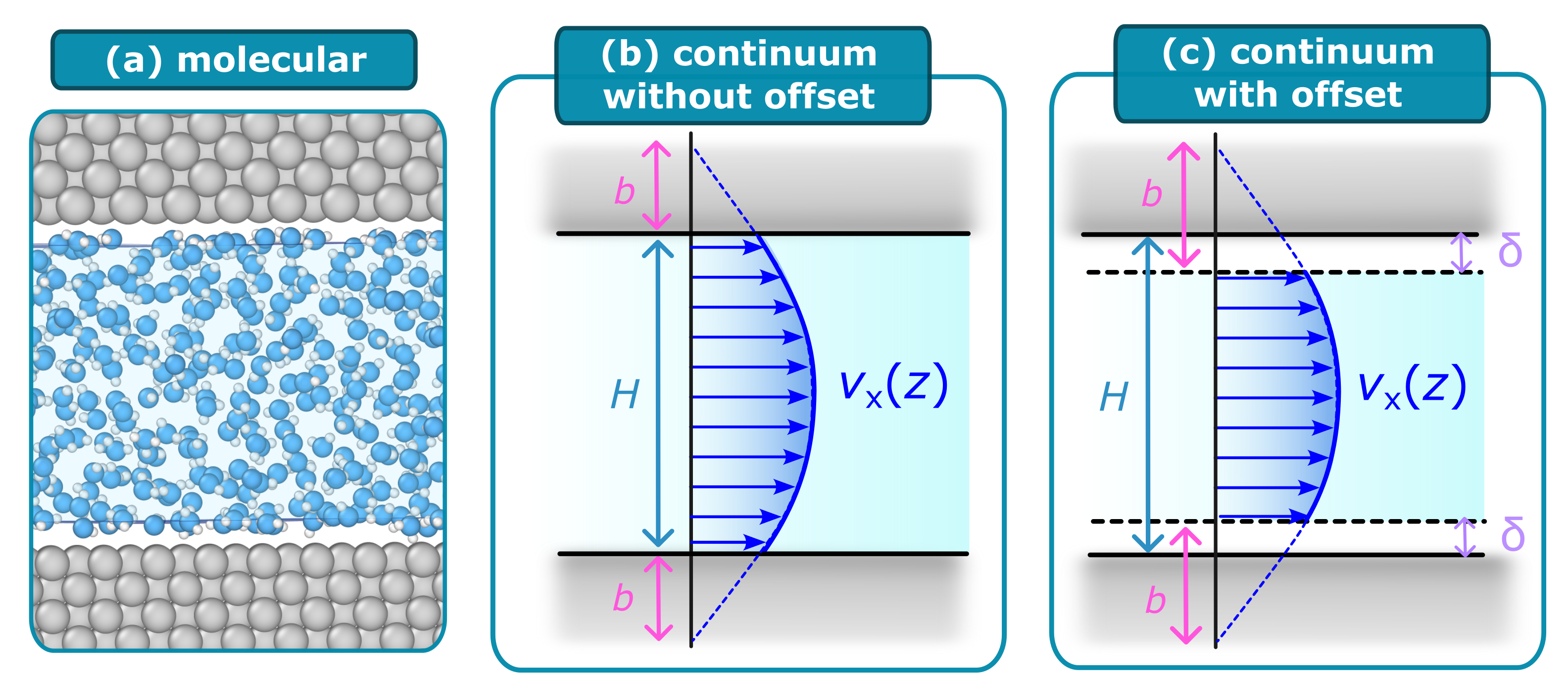}
  \caption{\textbf{Molecular and continuum representations of the system considered.}  (a) Water molecules (oxygen atoms in blue) are confined between two substrates with $\alpha=3$ (solid atoms in gray). (b) In the continuum representation without accounting for excluded volume at the interface, 
  the hydrodynamic boundary for each interface is place at the first plane of solid atoms in contact with the fluid. The fluid flow therefore spans across the whole channel height $H$. (c) More realistically, excluded volume can be accounting for by placing the hydrodynamic boundary away from the first atomic plane with an offset $\delta$.  The slip length $b$ is  the distance from the adjusted boundary to where the linearly extrapolated fluid velocity profile $v_x(z)$ vanishes.}
\label{fig:hydrodynamic_boundary}
\end{figure}

In writing Eq.~\ref{eqn:slip_length_tube} and Eq.~\ref{eqn:slip_length_sym},
we have made a simplifying approximation when 
treating the liquid as a continuum in CHT that 
the hydrodynamic boundary applies at the first plane of solid atoms in contact with the liquid.
Therefore, the solution for the velocity profile from
the Navier--Stokes equations span across the total height of the channel $H$ or the radius of the nanotube $R$.
However, microscopically, there is an offset
in the location of the hydrodynamic boundary due to
the exclude volume at each solid--liquid interface,
as schematically shown in Fig.~\ref{fig:hydrodynamic_boundary}.
To more realistically account for this excluded volume, 
the boundary conditions can be modified as follows
in the case of 1D Poiseuille flow
\begin{equation}
    \left.\frac{\partial v_z(r)}{\partial r}\right\rvert_{r=R-\delta}=-\frac{1}{b_R v_z(R-\delta)},
\end{equation}
where $\delta$ is the length
scale associated with the offset.
The resulting effective friction is therefore
\begin{equation}
    \lambda^{\mrm{1D}}_{\mrm{eff}} = \frac{4\eta}{R-\delta + 4b_R},
\end{equation}
and the curvature-dependent slip length becomes
\begin{equation}
    b_R = \frac{\eta}{\lambda^{\mrm{1D}}_{\mrm{eff}}} - \frac{R-\delta}{4}. 
\end{equation}
Similarly, in the case of 2D  Poiseuille flow, the modified boundary conditions are
\begin{equation}
    \left.\frac{\partial v_x(z)}{\partial z}\right\rvert_{z=\delta_1}=\frac{1}{b_1}v_x(\delta_1), \quad\quad \left.\frac{\partial v_x(z)}{\partial z}\right\rvert_{z=H-\delta_2}=-\frac{1}{b_2}v_x(H-\delta_2),
\end{equation}
where $\delta_1$ and $\delta_2$ are offsets for the hydrodynamic boundaries at the bottom and the top respectively.
The resulting effective friction for the general case of an asymmetric channel is therefore
\begin{equation}
    \lambda^{\mrm{2D}}_{\mrm{eff}} = \frac{12(H-\delta_1-\delta_2 + b_1 + b_2)\eta}{(H-\delta_1-\delta_2)^2 + 4 (H-\delta_1-\delta_2)(b_1 + b_2)+12b_1 b_2}.
    \label{eq:effective_friction_offset}
\end{equation}
As before, by considering a symmetric channel where 
$b_1=b_2=b$ and $\delta_1=\delta_2=\delta$,
the slip length is given as
\begin{equation}  
 b= \left(\frac{\eta}{\lambda^{\mrm{2D}}_{\mrm{eff}}} - \frac{H-2\delta}{3}\right) + \left[\left(\frac{\eta}{\lambda^{\mrm{2D}}_{\mrm{eff}}} - \frac{H-2\delta}{3}\right)^2 + (H-2\delta) \left(\frac{\eta}{\lambda^{\mrm{2D}}_{\mrm{eff}}} - \frac{H-2\delta}{12}\right)\right]^{1/2}.
 \label{eqn:slip_length_sym}
\end{equation}
There is an inherent ambiguity in the offset length scale since there is no simple microscopic expression for $\delta$ for each liquid--solid interface. In practice, we can provide a estimate
for $\delta$ from simulations to assess the sensitivity of
accounting for this offset, as will be discussed in in the next section.

\subsection{Results} \label{sec:sensitivity_boundary}

So far,
we have made a simplifying approximation when 
treating the liquid as a continuum in CHT that 
the hydrodynamic boundary applies at the first plane of solid atoms in contact with the liquid.
As discussed in Section~\ref{sec:sensitivity_theory}, it is more realistic that there is an offset $\delta$
as to where to hydrodynamic boundary is located
due to excluded volume at the liquid--solid interface.
In this section, we will examine the sensitivity of the choice of hydrodynamic boundary on the results.

\begin{figure}[H]     \centering
  \includegraphics[width=\linewidth]{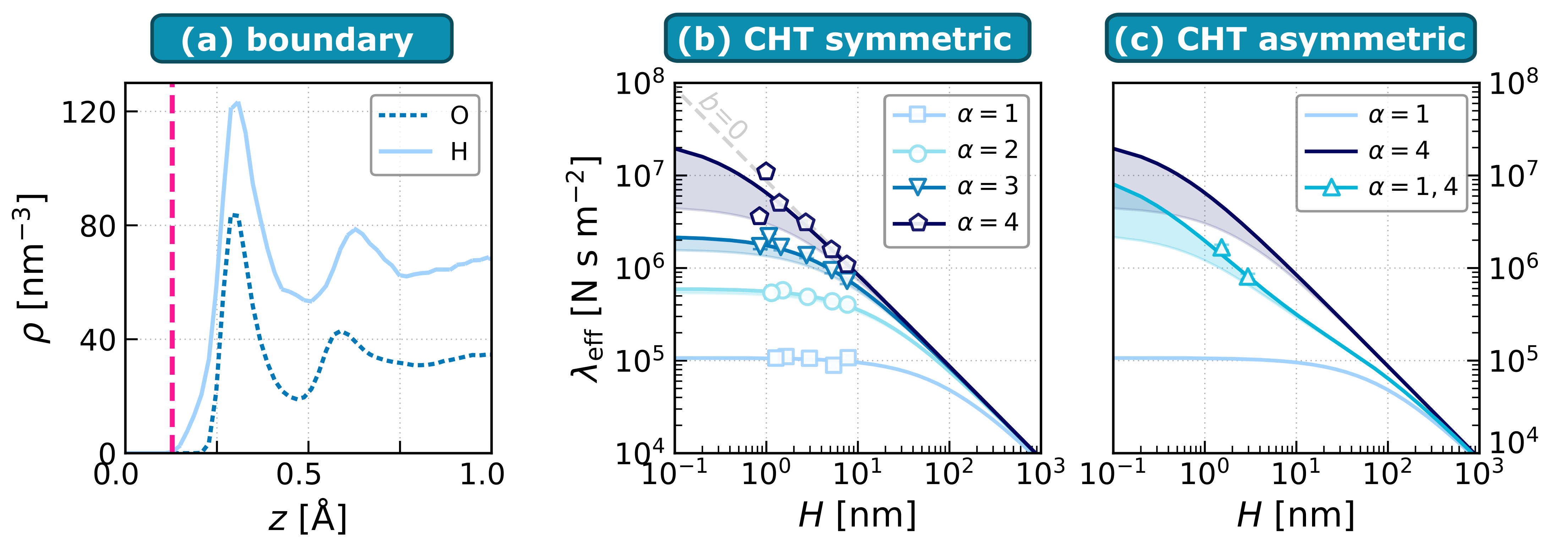}
  \caption{\textbf{The effective friction as a function of the channel height and its sensitivity to the hydrodynamic boundary position.}  
  (a) Number density profiles $\rho(z)$ for hydrogen (dark dotted line) and oxygen (light solid line) atoms of water near the 
  solid wall with $\alpha=3$ located at the $z=0$ plane. The dashed pink vertical line marks where the hydrogen atom density vanishes,
  giving an estimate of the offset between the hydrodynamic boundary and the solid wall $\delta=3.2\,\mrm{\AA}$.  The effective frictions
  $\lambda_{\mrm{eff}}$ predicted from CHT as a function of channel height $H$  are shown for (b) symmetric and (c) asymmetric channels. The solid lines are obtained 
  by $\delta=0$. The shaded region for each curve indicates the range of $\lambda_{\mrm{eff}}$ 
  obtained with $\delta=3.2\pm3.2\,\mrm{\AA}$,
   demonstrating the sensitivity of the result to
   the hydrodynamic boundary position. The data points indicated are obtained directly from EMD simulations using the Green--Kubo relation in Eq.~\ref{eq:GK_friction}. 
  }
\label{fig:boundary_sensitivity}
\end{figure}

In practice, we can estimate the lower limit of $\delta$ as the distance away from the 
first plane of solid atoms to the plane at which
the density of the fluid goes from zero to finite.
In Fig.~\ref{fig:boundary_sensitivity}(a), we marked this boundary for the $\alpha=3$ channel, where $\delta=3.2\,\mrm{\AA}$. 
Because the hydrogen atoms protrude further
toward the vapor phase than the oxygen atoms,
we have used the hydrogen density profile, as
done in Ref.~\onlinecite{Cox2022}.

In Figs.~\ref{fig:boundary_sensitivity}(b) and (c), we show the effective friction
$\lambda^{\mrm{2D}}_{\mrm{eff}}$
determined from the CHT solution that has accounted for an offset in Eq.~\ref{eq:effective_friction_offset}
as a function of $H$ for the different channels 
with its sensitivity to $\delta$. Here we have used
the slip lengths determined by the GK+CHT approach discussed in Section.~\ref{sec:sliplength}.
The effective friction $\lambda^{\mrm{2D}}_{\mrm{eff}}$ is the most sensitive to $\delta$
as the channel height decreases $H\to 0 $ and the slip length for the confining interface decreases
$b\to 0$. For the smallest channel with the lowest slippage, not accounting for the offset could lead
to a significant increase in the effective friction as the 
length scale of $H$ is comparable to that of $\delta$. Otherwise, in cases where $H>1\,\mrm{nm}$, then placing the hydrodynamic boundary at the first atomic plane of the solid is a reasonable approximation.
Simulation data deviating from the CHT prediction at the limit of sub-nanometric confinement will be further discussed in the next section.

\newpage

\section{Friction of water under confinement}

\subsection{System set-up}

\begin{figure}[H]     \centering
  \includegraphics[width=0.9\linewidth]{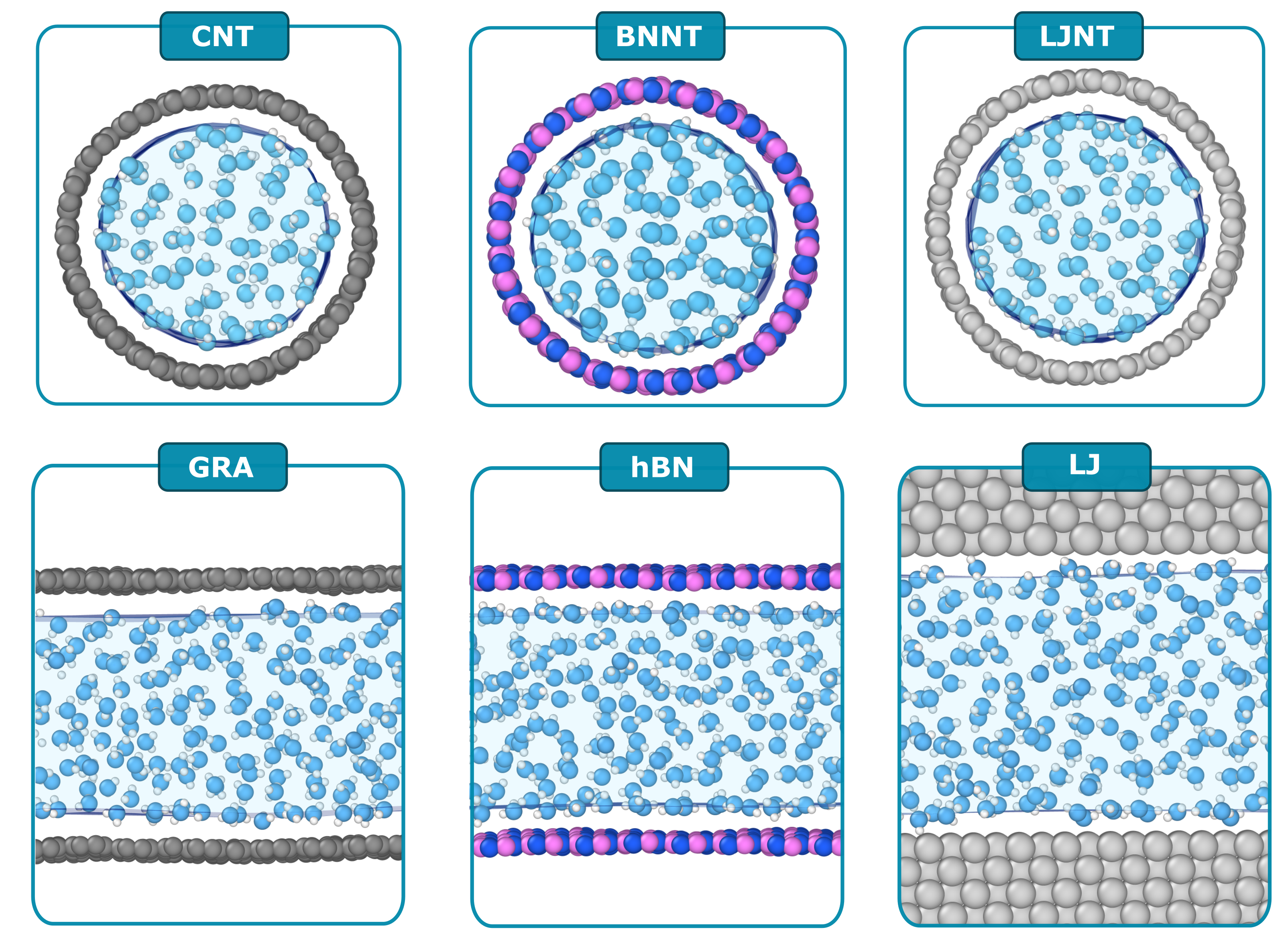}
  \caption{\textbf{Systems with water under 1D and 2D confinement  considered.}  Representative simulation snapshots are shown for the carbon nanotube (CNT), boron nitride nanotube (BNNT), Lennard--Jones nanotube (LJNT), graphene channel (GRA), hexagonal boron nitride channel (hBN) and channel made up two attractive Lennard--Jones walls with $\alpha=3$ (LJ).
  }
\label{fig:systems}
\end{figure}

The systems considered include: water confined in 1D nanotubes made up of different materials including carbon (CNT), boron nitride (BNNT) and Lennard--Jones particles (LJNT) and in 2D
slit made up of two sheets of graphene (GRA), hexagonal boron nitride (hBN) and two Lennard--Jones solid substrates with $\alpha=3$(LJ).
All nanotubes were of armchair
chirality $(m,m)$ where $m$ takes values between 12 and 40.  The CNTs and BNNTs considered include those considered in Ref.~\onlinecite{Thiemann2022}, with
the same simulation set-up.
The LJNTs considered has the same lattice as the CNTs of the same chirality.
The radii of the tubes considered range from $R\approx0.7\,\mrm{nm}$ to $R\approx2.8\,\mrm{nm}$.
The water density inside the nanotubes was set to the bulk limit of $33.36\,\mrm{nm}^{-3}$.
We report the system size and the number of water molecules for the nanotubes considered in Table.~\ref{tab:1Dsystems}.
For all the systems, the orthogonal simulation box with has lateral dimensions $\ell_z$ as the length of the nanotube, while
$\ell_x$ and $\ell_y$ are chosen to be large enough to minimise the tube interaction with its periodic images.

\begin{table}[H]
    \centering
    \begin{tabular}{c  c c  c c }
    \hline
    \hline
    System & $(m,m)$  & $R\,[\mrm{nm}]$  &  $\ell_z\,[\mrm{nm}]$ & $N$ \\
    \hline
    CNT & (40,40) & 2.72 & 2.96 &  2028  \\
    CNT & (35,35) & 2.38 & 2.96 &  1525  \\
    CNT & (30,30) & 2.04 & 2.96 &  1092  \\
    CNT & (25,25) & 1.70 & 2.96 &  731  \\
    CNT & (20,20) & 1.36 & 2.96 &  442  \\
    CNT & (18,18) & 1.23 & 2.96 &  347  \\
    CNT & (15,15) & 1.02 & 2.96 &  225  \\
    CNT & (12,12) & 0.82 & 2.96 &  130  \\
    CNT & (10,10) & 0.68 & 2.48 &  67 \\ 
    \hline
    BNNT & (40,40) & 2.77 & 3.01 &  2136  \\
    BNNT & (35,35) & 2.42 & 3.01 &  1607  \\
    BNNT & (30,30) & 2.08 & 3.01 &  1148  \\
    BNNT & (25,25) & 1.73 & 3.01 &  771  \\
    BNNT & (20,20) & 1.38 & 3.01 &  467  \\
    BNNT & (18,18) & 1.25 & 3.01 &  366  \\
    BNNT & (15,15) & 1.04 & 3.01 &  238  \\
    BNNT & (12,12) & 0.83 & 3.01 &  136  \\
    BNNT & (10,10) & 0.69 & 2.52 &  67 \\ 
    \hline
    LJNT & (40,40) & 2.77 & 2.96 &  2028  \\
    LJNT & (35,35) & 2.42 & 2.96 &  1525  \\
    LJNT & (30,30) & 2.08 & 2.96 &  1092  \\
    LJNT & (25,25) & 1.73 & 2.96 &  731  \\
    LJNT & (18,18) & 1.25 & 2.96 &  347  \\
    LJNT & (15,15) & 1.04 & 2.96 &  225  \\
    \hline
    \hline
    \end{tabular}
\caption
{
\textbf{Systems considered for 1D confinement of water.} For each system, we report the chirality of the single-walled nanotube $(m,m)$, the radius $R$, the length $L_z$ and the total number of confined water molecules $N$.}
\label{tab:1Dsystems}
\end{table}

For the case of 2D confinement, the channel heights considered range from $H\approx0.6\,\mrm{nm}$ to  $H\approx8.6\,\mrm{nm}$.
The GRA and hBN channels have AA stacking between the two sheets.
The LJ channel is made up of the same solid substrates considered in Section.~\ref{sec:verification}, with a wetting coefficint of $\alpha = 3$.
We report the system size and the number of water molecules for the channels considered in Table.~\ref{tab:2Dsystems}.
For all the systems, the orthogonal simulation box with has lateral dimensions $\ell_x \times \ell_y$ and $\ell_z$ is chosen to be large enough to minimise the slab interaction with its periodic images.

\begin{table}[H]
    \centering
    \begin{tabular}{c  c c  c  }
    \hline
    \hline
    System & $H\,[\mrm{nm}]$  & $\ell_x\times \ell_y\,[\mrm{nm}^2]$ & $N$ \\
    \hline
     GRA &   8.597 & 3.224 $\times$ 3.436 &  3080 \\ 
     GRA &   4.436 & 3.224 $\times$ 3.436 &  1540 \\ 
     GRA &   2.481 & 3.224 $\times$ 3.436 &  770 \\ 
     GRA &   1.505 & 3.224 $\times$ 3.436 &  409 \\ 
     GRA &   1.300 & 3.224 $\times$ 3.436 &  338 \\ 
     GRA &   1.078 & 3.224 $\times$ 3.436 &  258 \\ 
     GRA &   0.914 & 3.224 $\times$ 3.436 &  176 \\ 
     GRA &   0.672 & 3.224 $\times$ 3.436 &  87 \\
     \hline
     hBN &   8.597 & 3.278 $\times$ 3.494 &  3008 \\
     hBN &   4.436 & 3.278 $\times$ 3.494 &  1504 \\
     hBN &   2.457 & 3.278 $\times$ 3.494 &  752 \\
     hBN &   1.440 & 3.278 $\times$ 3.494 &  390 \\
     hBN &   1.277 & 3.278 $\times$ 3.494 &  328 \\
     hBN &   1.057 & 3.278 $\times$ 3.494 &  255 \\
     hBN &   0.930 & 3.278 $\times$ 3.494 &  166 \\
     hBN &   0.690 & 3.278 $\times$ 3.494 &  116 \\
     \hline
  LJ &   7.621 & 3.663 $\times$ 3.663 & 3240 \\
  LJ &   5.181 & 3.663 $\times$ 3.663 & 2160 \\
  LJ &   2.750 & 3.663 $\times$ 3.663 & 1080 \\
  LJ &   1.440 & 3.663 $\times$ 3.663 & 498 \\
  LJ &   1.064 & 3.663 $\times$ 3.663 & 332 \\
  LJ &   0.859 & 3.663 $\times$ 3.663 & 233 \\
    \hline
    \hline
    \end{tabular}
\caption
{
\textbf{Systems considered for 2D confinement of water.} For each system, we report the equilibrium height $H$, the lateral dimensions $\ell_x \times \ell_y$ and the total number of confined water molecules $N$.}
\label{tab:2Dsystems}
\end{table}

\subsection{Simulation details}

For simulations of water confined in CNTs, BNNTs, GRA and hBN channels, the interatomic interaction is described with density functional theory 
within the generalized gradient approximation using the dispersion-corrected functional revPBE-D3 \cite{Grimme2010,Zhang2010,Grimme2011}.
Since thorough sampling of the phase
space at this level of theory is prohibitively expensive, 
the machine-learned interatomic potentials (MLIPs) 
trained on forces and energies from ab initio simulation trajectories 
from Ref.~\onlinecite{Thiemann2022} were used.
EMD simulations were run in \texttt{LAMMPS} \cite{Thompson2022}
with an interface to the \texttt{n2p2} package\cite{Singraber2019}.
All atoms were treated as flexible. 
Moreover, deuterium masses were employed for the hydrogens to
ensure a stable simulations at computationally feasible time-steps. 
Dynamics were propagated using the velocity Verlet algorithm with time-step of 1 fs. 
The simulations were carried out in the NVT ensemble,
where the temperature was maintained at $T=300\,$K. 
Two separate stochastic velocity rescaling thermostats 
(CSVR) \cite{Bussi2007} were used, one applied to the liquid and one applied to the solid, both with a damping constant of 1 ps.
For simulations of the nanotubes, the linear momentum of the liquid was constrained at zero in the $z$ direction. 
For simulations of the channels, the linear momentum of the liquid was constrained at zero in the $x$ and $y$ directions.
In order to work in the frame of reference of the solid, the solid's center of mass velocity was also zeroed.
An
equilibration run of $100\,\mrm{ps}$ was performed, followed by production runs that are at least $5\,\mrm{ns}$ long.
During the production, the total summed force acting on all
water molecules in the $z$ direction was sampled 
at each time-step.

For simulations of water confined in LJNTs, 
water--water interactions were described with the SPC/E 
water model \cite{Berendsen1987}. 
The geometry of water molecules was constrained using 
the \texttt{RATTLE} algorithm \cite{Andersen1983}.
The atoms on the nanotubes were held rigid with respect to its lattice. Each oxygen atom on a water molecule interacts with
a substrate atom through a 12-6 Lennard-Jones potential 
with a well-depth of $\varepsilon_{\mrm{wf}}=0.3748\,\mrm{kcal\,mol^{-1}}$ and $\sigma_{\mrm{wf}}=0.3059\,\mrm{nm}$. This potential is
chosen based on water--graphite interaction strength, where the attraction strength
chosen $\varepsilon_{\mrm{wf}}$ was 4 times higher than the Werder parameters \cite{Werder2003}, in order to describe highly-wetting nanotubes.
All Lennard-Jones interactions were truncated and shifted at 
$1\,\mrm{nm}$. 
Electrostatic interactions were cut off at $1\,\mrm{nm}$ and 
long-ranged interactions were evaluated using particle--particle 
particle--mesh Ewald summation \cite{hockney1988} such that
the RMS error in the forces was a factor of $10^5$ smaller 
than the force between two unit charges separated by a 
distance of $0.1\,\mrm{nm}$ \cite{Kolafa1992}.
Dynamics were propagated using the velocity Verlet 
algorithm with 
time-step of $\SI{1}{\femto\second}$.
The simulations were carried out in the canonical 
(NVT) ensemble. The temperature of liquid water
is maintained at $T=298\,\mrm{K}$ using a
Nos\'{e}--Hoover chain with 5 thermostats and a damping constant of $\SI{0.1}{\pico\second}$.
For the constrained EMD simulations, the linear momentum of the liquid is zeroed
in the $z$ and $y$ directions by substracting the center-of-mass
velocity of the water at every time-step.
An
equilibration run of $200\,\mrm{s}$ was performed, followed by
a production run for each system that is $10\,\mrm{ns}$ long.
During the production, the total summed force acting on all
water molecules in the $z$ direction was sampled 
at each time-step.

For simulations of water confined in channels made with substrates of a high wetting coefficient $\alpha=3$, the details for the constrained EMD simulations
have been given in Section.~\ref{sec:verification}.

\subsection{Green--Kubo relation for the effective friction} 

\begin{figure}[H]     \centering
  \includegraphics[width=0.95\linewidth]{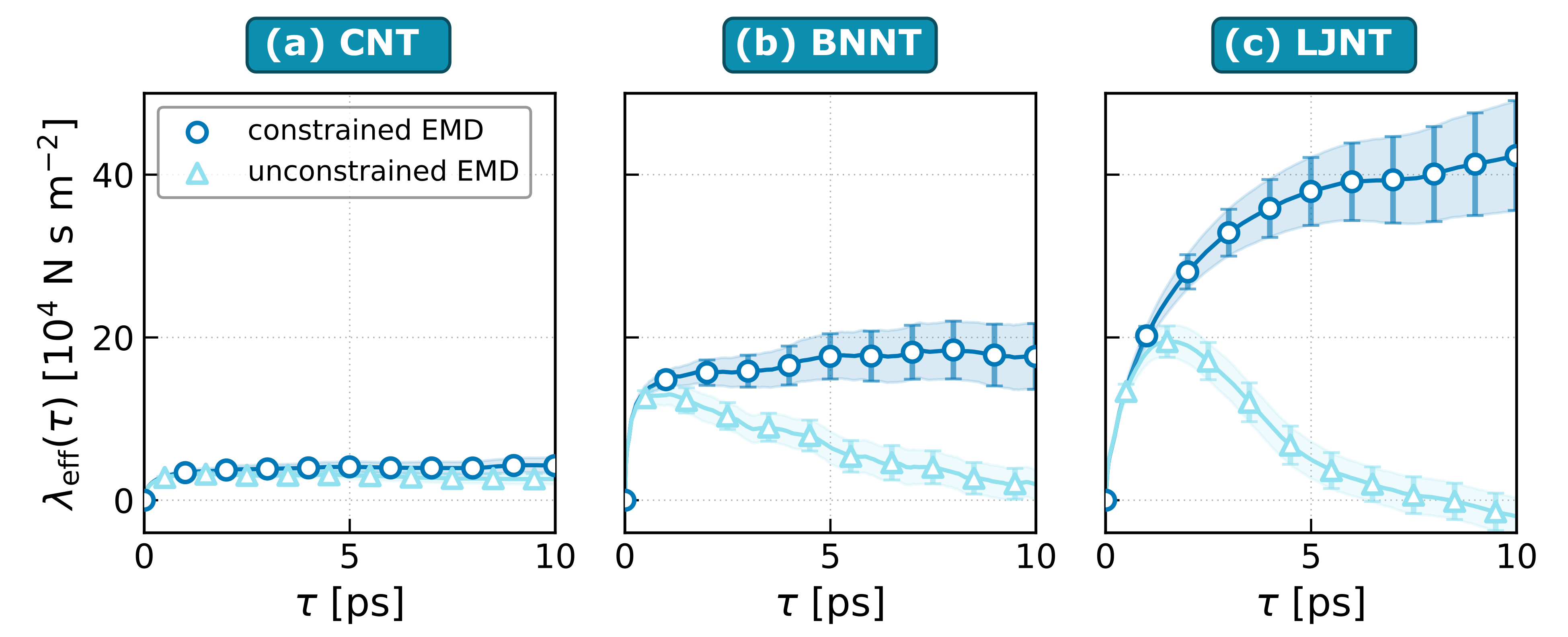}
  \caption{\textbf{Green--Kubo friction in 1D confinement.} The results
  are shown for a (a) CNT, (b) BNNT and (c) LJNT of the same chirality (20,20).
 In EMD simulations with the constraint
zeroing the liquid momentum along the tube direction, the Green–Kubo integral in Eq.~\ref{eq:GK_friction_1D} gives a well-defined plateau as a measurement of
the effective friction $\lambda^{\mrm{1D}}_{\mrm{eff}}$. When no constraint is applied, the integral decays to zero at long time. Such a plateau
problem is more severe for tubes with lower slip length.
   }
\label{fig:GK_friction_1D}
\end{figure}

For water flow inside the nanotube, the Green--Kubo relation for the effective friction is given as
\begin{equation}
    \lambda^{\mrm{1D}}_{\mrm{eff}} = \frac{\beta}{\mcl{A}}\int^\infty_0\!\mrm{d}\tau\,\langle  F_z(\tau) F_z(0)  \rangle_{\mrm{c}},
    \label{eq:GK_friction_1D}
\end{equation}
where the correlation is of the frictional force on the fluid along the direction of the tube and the appropriate constraint applied is on the momentum of the fluid along the tube 
\begin{equation}
    \dot{P}_z = 0, \quad\quad P_z = 0.
\end{equation}
The Green--Kubo friction computed from EMD simulations
as the integral of the force autocorrelation function
with and without the zero momentum constraint for the nanotubes
considered are shown in Fig.~\ref{fig:GK_friction_1D}. 
The error bars correspond to the statistical errors obtained from splitting the entire trajectory into 100 blocks such that each block is $100\,\mrm{ps}$ long.
As have discussed, while the Green--Kubo integral from the unconstrained EMD simulation decays to zero at longer time, it reaches
a well-defined plateau in the constrained EMD simulations,
which is the effective friction $\lambda^{\mrm{1D}}_{\mrm{eff}}$.

\subsection{Slip length and intrinsic friction prediction}

The slip length predictions using the GK+CHT method approach described in Section~\ref{sec:sliplength}
are shown in the Fig.~3(a) of the main paper.
We used an experimental \cite{Harris2004} value of $\eta=0.896\,\mrm{mPa\,s}$ for the shear viscosity of water as $\eta$ has not been computed
for water described with the functional. To obtain a smooth function for the curvature-dependent slip length $b_R(R)$, for each surface,  a function of the form
\begin{equation}
    b_R(R)= b_{\infty} + \frac{m}{R^n},
    \label{eq:fittingfunction}
\end{equation}
where $b_{\infty}$ is the slip length of the flat surface,
was used as a fitting function.
To obtain $b_{\infty}$, simulations of the 2D channel made up by the flat surfaces are used to obtained $\lambda^{\mrm{2D}}_{\mrm{eff}}$, which is then used in
Eq.~\ref{eqn:slip_length_sym} to obtain the slip length.
The optimal choices of $m$ and $n$
were determined using the True Region Reflective algorithm \cite{Branch1999}
as implemented in SciPy's \texttt{curve fit} routine \cite{Scipy2020} and are given in Table~\ref{tab:curvature_slip_length}.
The function $b_R(R)$ determined is then used in
the hydrodynamic solution in Eq.~\ref{eq:hydrodynamic_1D_eff}
to give the effective friction in 1D confinement as a function of radius as presented in the main paper.

\begin{table}[H]
    \centering
    \begin{tabular}{c  c c  c  }
    \hline
    \hline
    System & $m\,[\mrm{nm}^2]$  & $n$ & $b_\infty$ [nm] \\
    \hline
     CNT & 36.47 & 2.92 & 16.1  \\ 
     BNNT &  5.80 & 2.70 & 3.67  \\ 
    LJNT &  2.41 & 0.65 & 0.0  \\ 
    \hline
    \hline
    \end{tabular}
\caption
{
\textbf{Parameters for the curvature-dependent slip length.} When used in Eq.~\ref{eq:fittingfunction},
these parameters give the slip length of the nanotubes considered  as a function of the tube radius, determined by fitting to simulation data.
}
\label{tab:curvature_slip_length}
\end{table}

\begin{figure}[H]     \centering
  \includegraphics[width=0.95\linewidth]{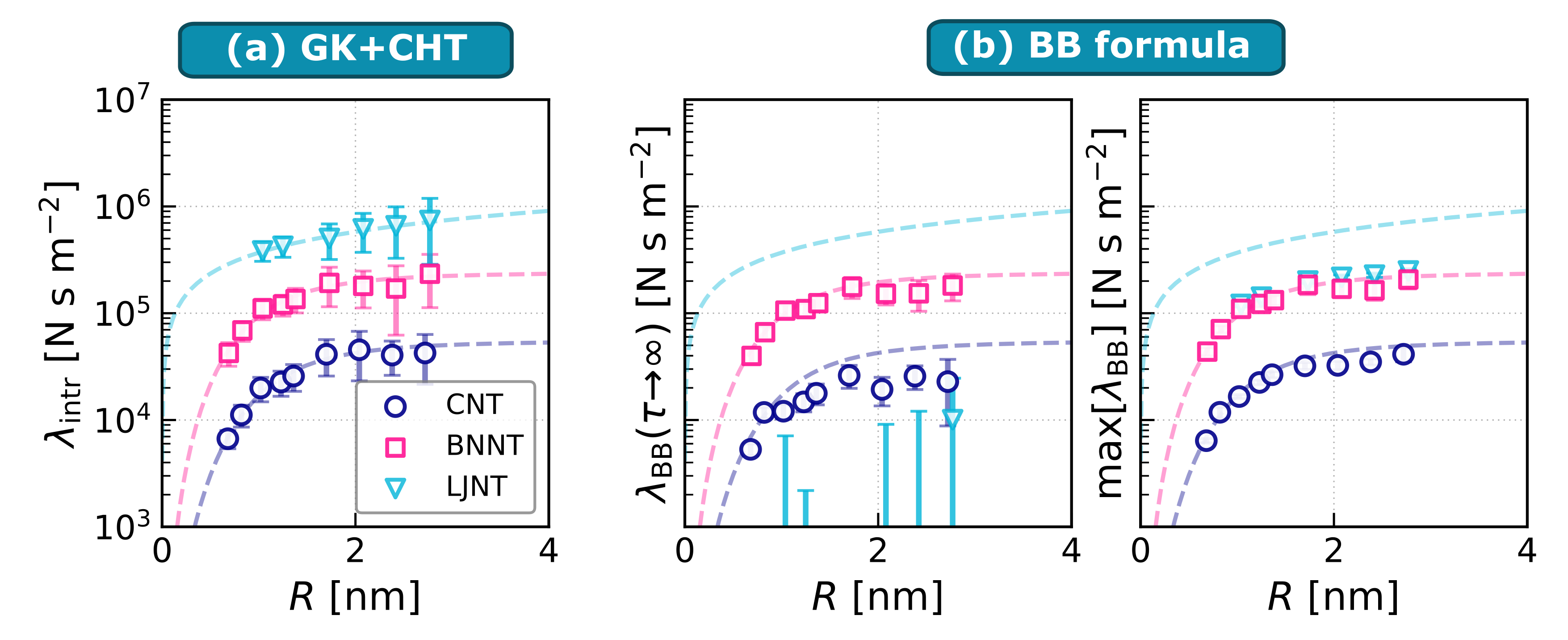}
  \caption{\textbf{The intrinsic friction of water in nanotubes.}
  Panel (a) shows the curvature dependence of the intrinsic friction coefficient for water on different surfaces obtained 
  from the GK+CHT method. The dashed lines are fits using Eq.~\ref{eq:fittingfunction} to simulation data from the GK+CHT method. Panel (b) shows results obtained from the BB formula by taking the long time limit (left) or taking the maximum value of the integral (right).
   }
\label{fig:curvature_dependent_intrinsic_friction}
\end{figure}

From the slip length, the intrinsic friction coefficient can be obtained from
\begin{equation}
    \lambda_{\mrm{intr}} = \frac{\eta}{b_R},
\end{equation}
which is shown in Fig.~\ref{fig:curvature_dependent_intrinsic_friction}.
For comparison, we also shown the results obtained by employing the BB formula, where the predictions become spurious for
surfaces of higher friction.

\subsection{Sub-nanometric confinement}

\begin{figure}[H]     \centering
  \includegraphics[width=\linewidth]{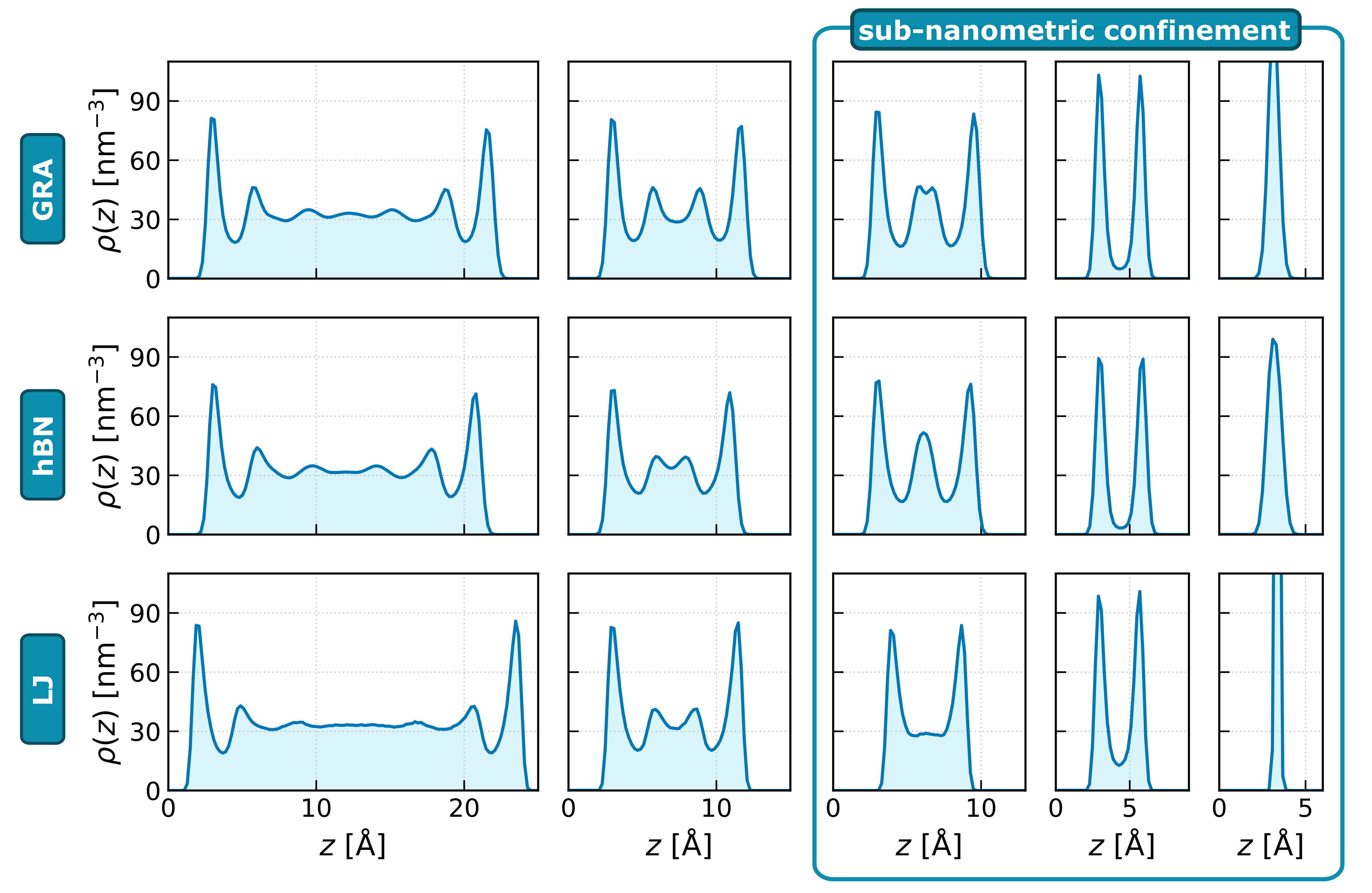}
  \caption{\textbf{Density profiles of water under 2D confinement.} The results are shown for the three 2D channels considered (GRA, hBN, LJ) of different heights. Overall, the structure of water remains
  broadly similar between the different confining materials. For the channels with $H\lesssim
 1\,\mrm{nm}$, it is no longer possible to have a well-defined bulk region. We denote this region as sub-nanometric confinement. }
\label{fig:density_confinement}
\end{figure}

\begin{figure}[H]     \centering
  \includegraphics[width=0.9\linewidth]{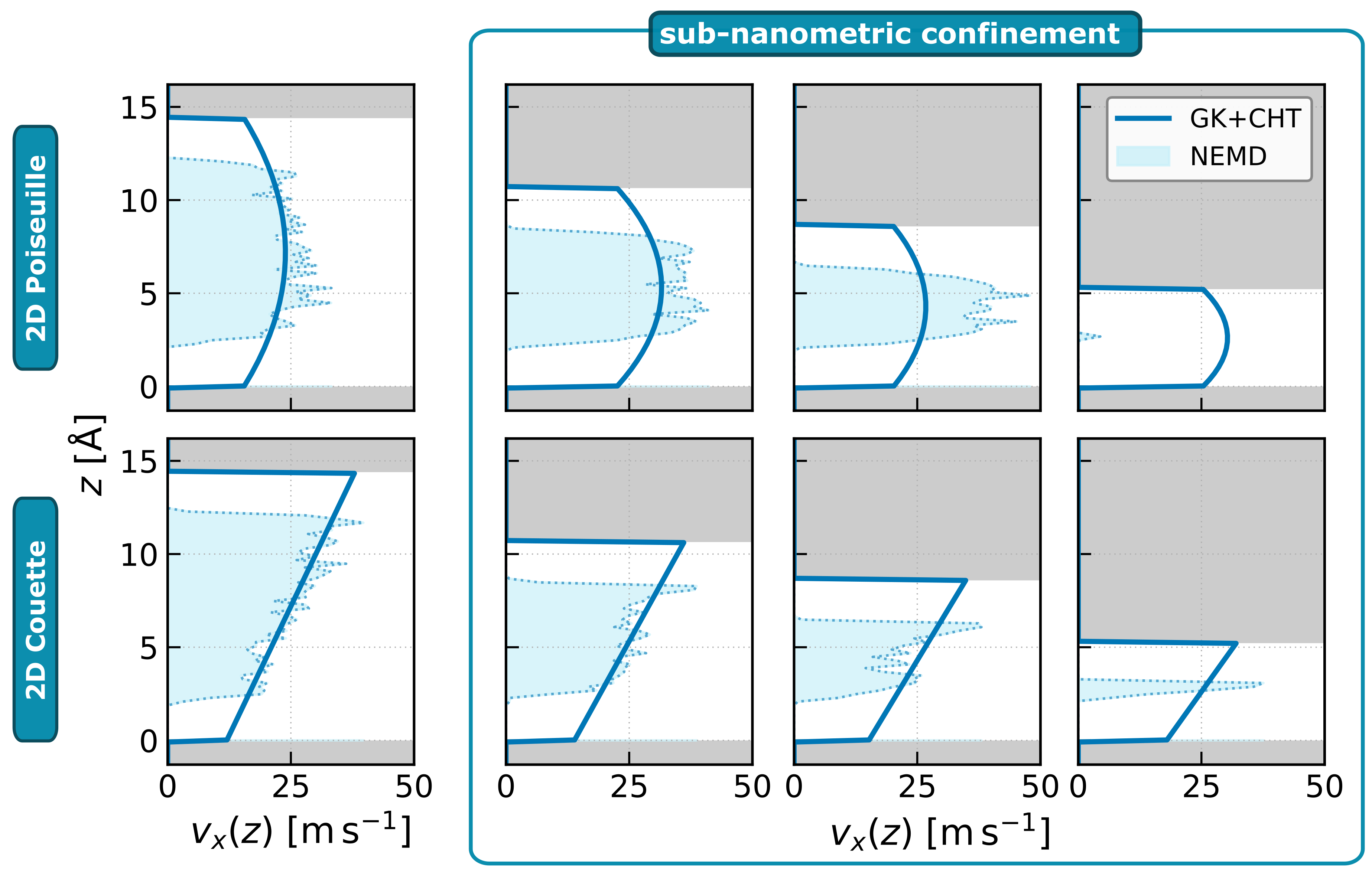}
  \caption{\textbf{Breakdown of CHT for water flow  under sub-nanometric confinement.}
   Comparison between the predicted velocity profiles from
classical hydrodynamic theory (CHT) with boundary determined with the Green--Kubo (GK) relation and direct NEMD
data are shown for 2D Poiseuille and 2D Couette flow. The deviation for the smallest of channels signifies the breakdown of CHT for channels too narrow for water to form a well-defined bulk region in the sub-nanometric confinement regime. }
\label{fig:ultraconfinement_LJ3}
\end{figure}

In cases of sub-nanometric confinement, 
we see that prediction of friction from
simulations starts to deviate from CHT.
This is because the water confined in between is no longer well
approximated as a continuum since the region of
inhomogeneity spans the whole channel, 
as seen in Fig.~\ref{fig:density_confinement}.
As discussed in Section.~\ref{sec:sensitivity_boundary}, when
$H\to 0$, sensitivity of the result of CHT to the choice of hydrodynamic boundary also increases significantly.
While the deviation of GK+CHT prediction from NEMD reference simulations for the velocity profiles, shown  for the case of LJ in Fig.~\ref{fig:ultraconfinement_LJ3}, is non-negligible under sub-nanometric confinement, 
it is not catastrophic apart from when there is only a single layer of water.

\bibliography{references}